\documentclass[12pt]{iopart}
\usepackage{iopams}
\usepackage{graphicx}
\usepackage[usenames,dvipsnames]{color}
\input{epsf}
\newcommand{\hyp}{\,_2F_1}

\begin{document}

\title[Diffusion with Gamma resetting]{Diffusion processes with Gamma-distributed resetting and non-instantaneous returns}

\author{Mattia Radice$^{1,2}$}
\address{$^1$ Dipartimento di Scienza e Alta Tecnologia and Center for Nonlinear and Complex Systems, Universit\`a degli Studi dell'Insubria, Via Valleggio 11, 22100 Como Italy}
\address{$^2$ I.N.F.N. Sezione di Milano, Via Celoria 16, 20133 Milano, Italy}
\eads{\mailto{mattia.radice@uninsubria.it}}
\vspace{10pt}

\begin{abstract}
We consider the dynamical evolution of a Brownian particle undergoing stochastic resetting, meaning that after random periods of time it is forced to return to the starting position. The intervals after which the random motion is stopped are drawn from a Gamma distribution of shape parameter $ \alpha $ and scale parameter $ r $, while the return motion is performed at constant velocity $ v $, so that the time cost for a reset is correlated to the last position occupied during the stochastic phase. We show that for any value of $ \alpha $ the process reaches a non-equilibrium steady state and unveil the dependence of the stationary distribution on $ v $. Interestingly, there is a single value of $ \alpha $ for which the steady state is unaffected by the return velocity. Furthermore, we consider the efficiency of the search process by computing explicitly the mean first passage time. All our findings are corroborated by numerical simulations.
\end{abstract}

\vspace{2pc}

\section{Introduction}
In the last years, resetting has been widely recognized as a simple yet effective mechanism to optimize diffusion and search processes. The effects of restart on stochastic dynamics have been considered in many branches of physics \cite{EvaMaj-2011,DurHenPar-2014,EvaMaj-2014,MajSabSch-2015,NagGup-2016,EvaMaj-2018Refractory,BasKunPal-2019,MerBoyMaj-2020,Bre-2021,Gra-2021,MajMouSabSch-2021,MerBoy-2021,SanDasNat-2021,SchBre-2021,SinSanIom-2021}, including quantum physics \cite{Nav-2018,PerCarMag-2021,WalBot-2021} and biophysics \cite{ReuUrbKla-2014,RotReuUrb-2015,RolLisSanGri-2016,RobHadUrb-2019}, but also in different fields such as computer science \cite{LubSinZuc-1993,MonZec-2002} and economics \cite{StoSanKoc-2021}, to cite a few examples - see \cite{EvaMajSch-2020} for a review or \cite{PalKosReu-2021} for a more recent introduction on the subject.

To formulate the problem in a simple way, let us consider a Brownian particle with diffusion coefficient $ D $ starting its motion at $ x(0)=x_0 $ and initially evolving according to the Langevin equation
\begin{equation}\label{key}
	\dot{x}(t)=\xi(t),
\end{equation}
where $ \xi(t) $ is a Gaussian white noise with zero mean and $ \langle\xi(t+t')\xi(t)\rangle =2D\delta(t')$. We suppose that given a constant rate $ r $, in a small time window of size $ \mathrm{d}t $ the particle has a probability $ r\mathrm{d}t $ of being immediately reset to the starting point, from which the process starts anew. Overall, the evolution of the position $ x(t) $ in the interval $ (t,t+\mathrm{d}t) $ is hence given by
\begin{equation}\label{eq:Langevin_reset}
	x(t+\mathrm{d}t)=
	\cases{x(t)+\xi(t)\mathrm{d}t & with probability $1-r\mathrm{d}t$\\
	x_0 & with probability $r\mathrm{d}t$.}
\end{equation}
The resetting mechanism drastically affects the features of the diffusion process. For example the repeated returns to the starting location, while keeping the system away from equilibrium by constantly restoring the initial condition, also prevent the diffusive spreading of the position. Hence, contrarily to the reset-free process, a non-equilibrium stationary state (NESS) is eventually reached. Furthermore, the first passage properties are also altered: while the mean first passage time (MFPT) to a threshold for a diffusive particle is infinite, it becomes finite in the presence of resetting and can be minimized by choosing a proper value of the resetting rate $ r^* $. These striking properties have led in the last years to an increasing interest in the subject and a huge amount of work has been dedicated to investigate the effects of resetting applied to different systems, such as geometric Brownian motion \cite{StoSanKoc-2021}, fractional Brownian motion \cite{WanCheKan-2021}, scaled Brownian motion \cite{BodCheSok-2019,BodCheSok-2019-Renew}, Continuous Time Random Walks \cite{MonVil-2013,MenCam-2016,Shk-2017,MonMasVil-2017,BodSok-2020-CTRWRes}, Lévy flights \cite{KusMajSabSch-2014,KusGod-2015,ZhoXuDen-2021} and the telegraphic process modelling run-and-tumble dynamics \cite{EvaMaj-2018,Mas-2019,RAD-2021}. For example, the properties of the NESS have been investigated for both closed and open quantum systems \cite{PerCarMag-2021,PerCarLes-2021,MagCarPer-2022}, and interestingly these studies are accompanied by the analysis of theoretical models and experimental realizations, involving for example ultracold bosonic atoms in a tilted optical lattice \cite{MukSenMaj-2018} and quantum Ising chains \cite{PerCarMag-2021}, where different kinds of resetting protocols are considered. Another experimental example has been implemented with silica microspheres driven by optical tweezers \cite{TalPalSek-2020,BesBovPet-2020}.

From the point of view of experiments, resetting poses several non-trivial difficulties, one of which regards the mechanism whereby the system is relocated to the desired initial condition. For example in \cite{TalPalSek-2020} the authors describe experimental realizations where resetting is performed by the action of a dynamical phase, consisting in return motions at constant velocity or at constant time. As discussed previously, the original model assumes that the system is relocated instantly, hence to take into account the time cost of the return phase generalizations of the aforementioned formulations are required. Different models have been proposed in the literature, where resetting is implemented, e.g., with the action of optical traps \cite{MerBoyMaj-2020,SanDasNat-2021,GupPlaKun-2021}, or where random refractory times preceding the restart are considered \cite{EvaMaj-2018Refractory,MasCamMen-2019,MasCamMen-2019JSTAT}. To take into account general spatiotemporal correlations, various works have followed the idea instead that returns should be performed according to a deterministic law \cite{ZhoXuDen-2021,RAD-2021,PalKusReu-2019,PalKusReu-2019PRE,MasCamMen-2019PRE,PalKusReu-2020,BodSok-2020-BrResI}.

Another line of research aimed at generalizing  the basic description of resetting regards the introduction of time-dependent resetting rates, viz., waiting time distributions between resetting event different from the exponential distribution $ \psi(\tau)=r\exp(-r\tau) $. The domain of waiting time distributions considered so far includes Gamma and Weibull distributions \cite{EulMet-2016}, power-laws \cite{NagGup-2016,BodSok-2020-CTRWRes} and delta distributions, i.e., resetting at fixed times, which has been recognized as the most effective resetting protocol for search processes \cite{EulMet-2016,PalKunEva-2016,PalReu-2017,CheSok-2018}. In \cite{PalKunEva-2016} the authors consider the effects of a general $ r(t) $, providing a sufficient condition for the existence of a NESS and general formulae for the MFPT.

The scope of this paper is to investigate the combination of non-instantaneous returns and time-dependent resetting rates. In particular, we focus on resetting periods ruled by Gamma distributions and returns performed at constant velocity. The choice of the Gamma distribution represents a simple yet non-trivial example of a nonconstant rate whereby we are able to obtain explicit results, even when returns are non-instantaneous. Moreover, there are several results in the literature which show that for systems with resetting at constant rate and returns at constant velocity, the probability distribution of the position is completely unaffected by the return phase and coincides with that of the corresponding systems with instantaneous returns \cite{PalKusReu-2019PRE,PalKusReu-2019,BodSok-2020-BrResI,RAD-2021}. By using the Gamma distribution, we are able to test the validity of this result for nonconstant resetting rates: we will show indeed that in general the systems exhibits a velocity-dependent NESS and the independence of the return velocity is achieved only for exponentially distributed resetting times, namely when resetting follows Poissonian statistics.

The paper is organised as follows: in section \ref{s:model} we present the model and the general theory, which follows the same line of \cite{BodSok-2020-BrResI}. In sections \ref{s:Pst_calc}  and \ref{s:Pst_vdep} we study the NESS by evaluating the stationary distribution, and show that in general it depends explicitly on the value of the return velocity. In sections \ref{s:MFPT} and \ref{s:MFPT_halfInt} we investigate the first passage properties of the system and their dependence on the return dynamics, by providing explicit expressions for the MFPT. Finally, in section \ref{s:Concl} we draw our conclusions and summarize the results. 

\section{The model}\label{s:model}
In order to describe the model we follow the approach of \cite{BodSok-2020-BrResI}. The process is defined as a sequence of subprocesses. Each subprocess consists of two phases: the stochastic dynamics of the diffusing particle, that we call \textit{the displacement phase}, and the deterministic motion to the resetting location, which occurs after the resetting event and we call \textit{the return phase}. In the following we will always take $ x=0 $ as both the starting point of the motion and the restart location. Consider a subprocess starting at $ t=t_0 $. During the displacement phase the evolution of the probability distribution of the position is governed by the diffusion equation
\begin{equation}\label{key}
	\frac{\partial p(x,t)}{\partial t}=D\frac{\partial^2 p(x,t)}{\partial x^2},
\end{equation}
with solution relative to the initial condition $ p(x,t_0)=\delta(x) $:
\begin{equation}\label{eq:Diff_sol}
	p(x,t)=\frac{1}{\sqrt{4\pi D(t-t_0)}}\exp\left[-\frac{x^2}{4D(t-t_0)}\right].
\end{equation}
The duration $ \tau $ of the displacement phase corresponds to the time of the resetting event measured from $ t_0 $ and it is considered a random variable drawn from a Gamma distribution with density $ \psi(\tau) $. After a time $ \tau $ the particle occupies a random position $ x_0 $ from which the deterministic motion of the return phase starts. The evolution then follows the equation $ x(t)=\chi(t,x_0) $ and the time cost $ \theta(x_0) $ to perform the reset is given by the condition $ 0=\chi(\theta,x_0) $. In this phase the particle moves at constant velocity and is always directed towards the origin, therefore the equation of motion is
\begin{equation}\label{eq:law_of_motion}
	x(t)=\chi(t,x_0)=-\mathrm{sgn}(x_0)vt+x_0,
\end{equation}
where $ \mathrm{sgn}(y) $ denotes the sign of the argument and $ v>0 $ is the absolute value of the velocity. The time needed to return to the origin is thus
\begin{equation}\label{eq:time_cost}
	\theta(x_0)=\frac{|x_0|}{v}.
\end{equation}

\subsection{Duration of a subprocess}
The total duration of a subprocess is simply the sum of the durations of the displacement and return phases. The former corresponds to the random variable $ \tau $ and hence is distributed according to $ \psi(\tau) $; the latter is instead a function of the position occupied at the time of the resetting event, whose distribution is $ p(x_0,\tau) $. The total duration is thus $ t=\tau+|x_0|/v $ and by averaging over all possible values of $ \tau $ and $ x_0 $ we get
\begin{equation}\label{eq:phi}
	\phi(t)=\int_{0}^{\infty}\mathrm{d}\tau\psi(\tau)\int_{-\infty}^{+\infty}\mathrm{d}x\delta\left(t-\tau-|x|/v\right)p(x,\tau),
\end{equation}
where $ \delta(y) $ denotes the Dirac delta function. This equation can be recast more conveniently in Laplace space:
\begin{equation}\label{eq:phi_LT}
	\hat{\phi}(s)=\int_{-\infty}^{+\infty}\mathrm{d}x\rme^{-s|x|/v}\int_{0}^{\infty}\mathrm{d}\tau \rme^{-s\tau}\psi(\tau)p(x,\tau).
\end{equation}
The probability density $\phi_n(t)$ that the $ n $-th subprocess starts at time $ t $ is given by
\begin{equation}\label{eq:phi_n}
	\phi_n(t)=\cases{
	\delta(t) & for $n=1$\\
	\int_{0}^{t}\phi_{n-1}(t')\phi(t-t')\mathrm{d}t' & for $n\geq 2$,}
\end{equation}
where the delta function in the case $ n=1 $ accounts for the fact that the first subprocess starts at time $ t=0 $. By the convolution theorem, the Laplace transform of \eref{eq:phi_n} is simply
\begin{equation}\label{eq:phi_n_LT}
	\hat{\phi}_n(s)=\left[\hat{\phi}(s)\right]^{n-1},\quad n\geq1.
\end{equation}
The sum of all $ \phi_n(t) $ yields the renewal rate $ \kappa(t) $:
\begin{equation}\label{key}
	\kappa(t)=\sum_{n=1}^{\infty}\phi_n(t),
\end{equation}
which, from \eref{eq:phi_n_LT}, is expressed in Laplace domain as
\begin{equation}\label{eq:kappa_LT}
	\hat{\kappa}(s)=\frac{1}{1-\hat{\phi}(s)}.
\end{equation}
Note that for distributions possessing a well-defined first moment $ \eta $, in the small-$ s $ limit we can write $ \hat{\phi}(s)\sim1-\eta s $. Thus in the long time limit \eref{eq:kappa_LT} yields $ \hat{\kappa}(s)\sim 1/(\eta s) $, where $ \eta $ is the mean duration of a subprocess, and the renewal rate converges to a constant value:
\begin{equation}\label{key}
	\lim_{t\to\infty}\kappa(t)=\frac{1}{\eta}.
\end{equation}

\subsection{Probability density function and stationary distribution}
Let us first consider a single subprocess and, without loss of generality, suppose that it starts at $ t_0=0 $. Let $ G(x,t) $ be the probability density function (PDF) of the position for a subprocess. This quantity can be written as the sum of two terms, depending on the time of the resetting event. The first is the contribution of those walks that have not been reset up to time $ t $: in this case the displacement density is $ p(x,t) $, therefore the first term reads
\begin{equation}\label{eq:G_1}
	G_1(x,t)=p(x,t)\Psi(t),
\end{equation}
where $ \Psi(t) $ denotes the probability that no resetting event has occurred up to time $ t $:
\begin{equation}\label{key}
	\Psi(t)=\int_{t}^{\infty}\psi(t)\mathrm{d}t.
\end{equation}
The second term comes from the walks with resetting time $ \tau<t $: the displacement phase stops at a random location $ x_0 $ with distribution $ p(x_0,\tau) $, from which the position starts evolving according to the deterministic law of motion $ x=\chi\left(t,x_0\right) $, hence we can write
\begin{equation}\label{eq:G_2}
	G_2(x,t)=\int_{0}^{\infty}\mathrm{d}\tau\psi(\tau)\int_{-\infty}^{+\infty}\mathrm{d}x_0p(x_0,\tau)g_2(x,t;x_0,\tau),
\end{equation}
where $g_2(x,t;x_0,\tau)$ is
\begin{equation}\label{eq:g2}
	g_2(x,t;x_0,\tau) = \delta \left[x-\chi(t-\tau,x_0)\right]\Theta(t-\tau)\Theta(\tau+|x_0|/v-t),
\end{equation}
and $ \Theta(x) $ is defined as follows:
\begin{equation}\label{key}
	\Theta(x)=\cases{
		1 & if $x\geq 0$\\
		0 & if $x<0$.}
\end{equation}
The delta function in \eref{eq:g2} accounts for the fact that the position $x$ is determined by the law of motion, while the Theta functions ensure that the resetting happens before the observation time $ t $ (first Theta function) and that we are observing the system before the end of the subprocess (second Theta function).
	
Now we denote with $ P(x,t) $ the PDF of the position for the complete process and observe that the probability of occupying position $ x $ at time $ t $ corresponds to the probability that the last subprocess starts at $ t'<t $ and then the particle reaches $ x $ in a time $ t-t' $. By integrating over all possible values of $ t' $ we get
\begin{equation}\label{eq:P_complete}
	P(x,t)=\int_{0}^{t}\kappa(t')G(x,t-t')\mathrm{d}t',
\end{equation}
which is more easily evaluated in Laplace space:
\begin{equation}\label{eq:P_complete_LT}
	\hat{P}(x,s)=\hat{\kappa}(s)\hat{G}(x,s)=\frac{\hat{G}(x,s)}{1-\hat{\phi}(s)},
\end{equation}
where we used the expression of $ \hat{\kappa}(s) $ in \eref{eq:kappa_LT} and $ \hat{G}(x,s)=\hat{G}_1(x,s)+\hat{G}_2(x,s) $. The contributions $ \hat{G}_1(x,s) $ and $ \hat{G}_2(x,s) $ can be deduced from \eref{eq:G_1} and \eref{eq:G_2}, respectively. The first is simply written as
\begin{equation}\label{eq:G_1_LT}
	\hat{G}_1(x,s)=\int_{0}^{\infty}\rme^{-st}p(x,t)\Psi(t)\rmd t,
\end{equation}
while for $ \hat{G}_2(x,s) $ we can change the order of integration between $ t $ and $ \tau $, and by introducing $ u=t-\tau $ we arrive at
\begin{equation}\label{eq:G_2_LT}
	\fl \hat{G}_2(x,s)=\int_{0}^{\infty}\mathrm{d}\tau\psi(\tau)\int_{-\infty}^{+\infty}\mathrm{d}x_0p(x_0,\tau)\int_{0}^{\infty}\rmd u\rme^{-su}\delta\left[x-\chi(u,x_0)\right]\Theta(|x_0|/v-u),
\end{equation}
where the lower bound of integration in $ u $ follows from the Theta function $ \Theta(t-\tau) $ in the definition of $ g_2(x,t;x_0,\tau) $, see \eref{eq:g2}. We now observe that the integration variable $ u $ represents the time of the return phase, during which the distance of the particle from the origin monotonically decreases to zero. This means that the delta function in the last integral provides a contribution only for distances smaller than the initial distance, i.e., for $ |x|\leq|x_0| $. By considering the change of variable $ y=\chi(u,x_0) $, we can invert the equation of motion to write $ u=\vartheta(y,x_0) $ and thus the integral in $ u $ yields
\begin{equation}\label{key}
	\int_{0}^{|x_0|/v}\rmd u\rme^{-su}\delta\left[x-\chi(u,x_0)\right]=\frac 1v\Theta\left(|x_0|-|x|\right) \rme^{-s\vartheta(x,x_0)}.
\end{equation}
By plugging this expression in \eref{eq:G_2_LT} and exploiting the symmetry of $ p(x,t) $, we can finally write
\begin{equation}\label{key}
	\fl
	\hat{G}_2(x,s)=\frac1v\int_{0}^{\infty}\mathrm{d}\tau\psi(\tau)\int_{|x|}^{+\infty}\mathrm{d}x_0p(x_0,\tau)\times\cases{
	\rme^{-s\vartheta(x,x_0)} & for $x\geq0$\\
	\rme^{-s\vartheta(x,-x_0)} & for $x<0$.
}
\end{equation}
	
We can deduce the long time properties of $ P(x,t) $ by evaluating the small-$ s $ behaviour of $ \hat{P}(x,s) $. We assume that $ \phi(t) $ has finite mean $ \eta $, in which case in the small-$ s $ limit $ \hat{\phi}(s)\sim 1-\eta s $. Hence from \eref{eq:P_complete_LT} we can write
\begin{equation}\label{key}
	\hat{P}(x,s)\sim \frac{1}{\eta s}\hat{G}(x,s)\vert_{s=0}
	=\frac{1}{\eta s}\int_{0}^{\infty}G(x,t)\mathrm{d}t,
\end{equation}
whereby one can deduce the following relation in the time domain:
\begin{equation}\label{eq:P_st_def}
	\lim_{t\to\infty}P(x,t)=P(x)=\frac 1\eta\left[\rho_1(x)+\rho_2(x)\right],
\end{equation}
with
\begin{eqnarray}\label{key}
	\rho_1(x)&=\int_{0}^{\infty}G_1(x,t)\mathrm{d}t\\
	\rho_2(x)&=\int_{0}^{\infty}G_2(x,t)\mathrm{d}t.
\end{eqnarray}
Equation \eref{eq:P_st_def} defines the stationary distribution $ P(x) $, which in this case is the sum of two different contributions that can be computed as the limit $ s\to0 $ of $ \hat{G}_1(x,s) $ and $ \hat{G}_2(x,s) $, yielding respectively
\begin{equation}\label{key}
	\rho_1(x)=\int_{0}^{\infty}p(x,t)\Psi(t)\rmd t,
\end{equation}
and
\begin{equation}\label{key}
	\rho_2(x)=\frac 1v\int_{0}^{\infty}\mathrm{d}\tau \psi(\tau)\int_{|x|}^{+\infty}\rmd x_0 p(x_0,\tau).
\end{equation}
Note that in principle one can derive the full time-dependent PDF from \eref{eq:P_complete_LT} by means of Laplace inversion, see for example \cite{PalKusReu-2019PRE} in the case of Poissonian resetting. We point out however that this task is not easy for more general types of waiting time distributions and goes beyond the scope of this paper.

\section{Computation of the stationary distribution}\label{s:Pst_calc}
\begin{figure}[h!]
	\centering
	\includegraphics[width=.55\linewidth]{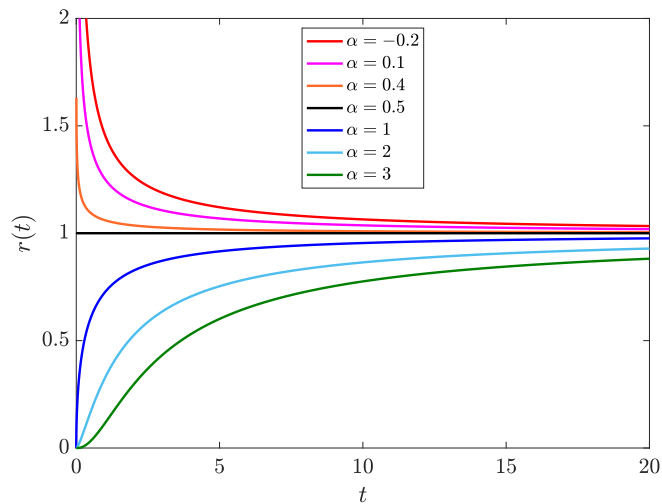}
	\caption{Rate function $ r(t) $ of the Gamma distribution given by \eref{eq:Gamma_rate}, for several values of the shape parameter $ \alpha $. The scale parameter $ r $ is set to unity. Note that for $ \alpha=0.5 $ one obtains a constant rate $ r(t)=r $.}
	\label{fig:rates}
\end{figure}

We are now ready to characterize the steady state by evaluating explicitly the long time limit of the PDF. We recall that we consider waiting times for the resetting events drawn from a Gamma distribution, with density
\begin{equation}\label{eq:Gamma_wait}
	\psi(\tau)=\frac{r\rme^{-r\tau}}{\Gamma\left(\alpha+\frac 12\right)}(r\tau)^{\alpha-\frac 12},\quad\alpha>-\frac 12,\,r>0.
\end{equation}
Here $ r $ is called \textit{scale parameter} and $ \alpha $ is called \textit{shape parameter}. Note that in the literature the shape parameter is often defined as $ \nu=\alpha+\case{1}{2} $, we are using this notation for the sake of clarity in the following expressions. It follows from \eref{eq:Gamma_wait} that the probability of not observing any event up to time $ t $ is given by
\begin{equation}\label{eq:Gamma_surv}
	\Psi(t)=\int_{t}^{\infty}\psi(\tau)\mathrm{d}\tau=\frac{\Gamma\left(\alpha+\frac12,rt\right)}{\Gamma(\alpha+\frac 12)},
\end{equation}
where $ \Gamma(\nu,z) $ is an upper incomplete gamma function \cite{Abr-Steg}. According to our notation, $ \alpha=\case{1}{2} $ corresponds to an exponential distribution with scale $ r $, which represents the case of events (resetting) occurring at constant rate $ r $. However, in general the Gamma distributions have a time-dependent rate $ r(t) $. Indeed, the rate function $ r(t) $ is defined as \cite{Ros}
\begin{equation}\label{key}
	r(t)=\frac{\psi(t)}{\Psi(t)},
\end{equation}
which can be interpreted as the conditional probability density that an event which has not occurred up to time $ t $ will occur in the next moment. From \eref{eq:Gamma_wait} and \eref{eq:Gamma_surv} one obtains the expression
\begin{equation}\label{eq:Gamma_rate}
	r(t)=\frac{r\mathrm{e}^{-rt}(rt)^{\alpha-\frac 12}}{\Gamma\left(\alpha+\frac 12,rt\right)},
\end{equation}
and it can be verified that $ r(t) $ is increasing for $ \alpha>\case 12 $, decreasing for $ \alpha<\case 12 $ and constant for $ \alpha=\case 12 $, furthermore for any $ \alpha $ it converges in the long time limit to the fixed value $ r $, see figure \ref{fig:rates} for a few examples. Hence, the scale parameter of the Gamma distribution controls the asymptotic value of the rate, while the shape parameter controls its time dependence.

The first step to compute the stationary distribution is the evaluation of the subprocess mean duration $ \eta $, which can be obtained from the Laplace transform $ \hat{\phi}(s) $ as
\begin{equation}\label{key}
	\eta=-\left.\frac{\rmd \hat{\phi}(s)}{\rmd s}\right\vert_{s=0}.
\end{equation}
To compute $ \hat{\phi}(s) $, we consider \eref{eq:phi_LT} and integrate first over $ \tau $ and then over $ x $. By using the results \eref{eq:app:I1} and \eref{eq:app:I2} in \ref{app:Integrals} we arrive at
\begin{equation}\label{eq:phi_LT_expression}
	\fl\hat{\phi}(s)=\frac{2\Gamma(2\alpha+1)}{\Gamma\left(\alpha+\frac 12\right)\Gamma\left(\alpha+\frac 32\right)}\left(\frac{r}{r+s}\right)^{\alpha+\frac 12}\frac{\hyp\left(2\alpha+1,\alpha+\frac 12;\alpha+\frac 32;1-\frac{2}{\zeta}\right)}{\zeta^{2\alpha+1}},
\end{equation}
where $ \hyp(a,b;c;z) $ is a Gauss hypergeometric function \cite{Abr-Steg} and the variable $ \zeta $ is defined as
\begin{equation}\label{key}
	\zeta=1+\frac{s}{v}\sqrt{\frac{D}{r+s}}.
\end{equation}
By taking the derivative of \eref{eq:phi_LT_expression} one can verify that the mean duration of a subprocess is
\begin{equation}\label{eq:eta}
	\eta_\alpha=\frac{\alpha+1/2}{r}+\frac{C_\alpha}{v}\sqrt{\frac{D}{r}},
\end{equation}
where $ C_{\alpha} $ is a constant given by
\begin{equation}\label{eq:C_alpha}
	C_{\alpha}=\frac{2}{\sqrt{\pi}}\frac{\Gamma(\alpha+1)}{\Gamma\left(\alpha+\frac 12\right)}.
\end{equation}
The result of \eref{eq:eta} is easy to interpret: the first term on the right-hand side is the mean duration of the displacement phase, as one can verify by computing the first moment of $ \psi(\tau) $; the second term instead represents the average return time, which can be written as $ \langle x_0\rangle/v$, with $ \langle x_0\rangle $ denoting the mean position occupied at the moment of the resetting event.

\begin{figure}[h!]
	\centering
	\includegraphics[width=.55\linewidth]{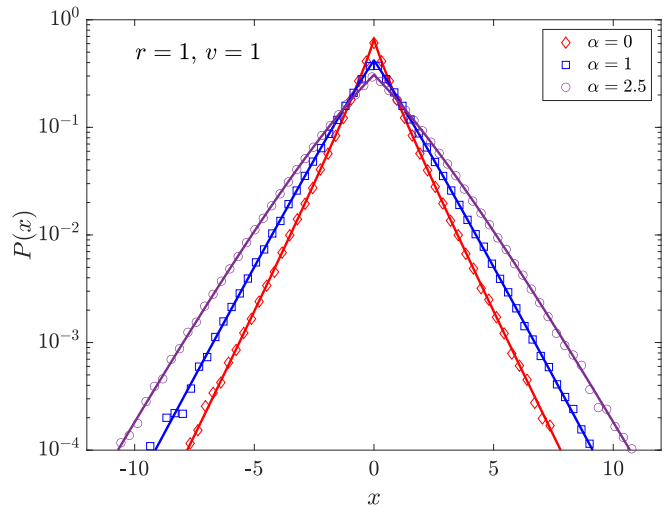}
	\caption{Stationary distribution for $ \alpha=0$ (red diamonds), $\alpha=1$ (blue squares) and $\alpha=2.5 $ (purple circles). In each case, data are obtained by simulating $ 10^6 $ walks with time step $ \rmd t=0.01 $ up to time $ t=10 $, and compared to the corresponding $ P(x) $ given by \eref{eq:P_st_expression}. The values of the return velocity, the scale parameter and the diffusion coefficient are set to unity.}
	\label{fig:stat}
\end{figure}

We can now proceed with the computation of $ \rho_1(x) $ and $ \rho_2(x) $. The first term may be written as
\begin{eqnarray}
	\rho_1(x)&=\int_{0}^{\infty}\rmd tp(x,t)\int_{t}^{\infty}\rmd \tau\psi(\tau)\\
	&=\int_{0}^{\tau}\rmd tp(x,t)\int_{0}^{\infty}\rmd \tau\psi(\tau)\\
	&=\int_{0}^{\infty}\rmd \tau\psi(\tau)\mathcal{I}(x,\tau),\label{eq:rho_1_nested}
\end{eqnarray}
with $ \mathcal{I}(x,\tau) $ defined by
\begin{equation}\label{eq:I}
	\mathcal{I}(x,\tau)=\int_{0}^{\tau}p(x,t)\rmd t,
\end{equation}
whose solution is given in \ref{app:Integrals} by formula \eref{eq:app:I3}. The integration over $ \tau $ finally yields an explicit expression for $ \rho_1(x) $, see \eref{eq:app:I4}. Let us introduce
\begin{equation}\label{key}
	z=|x|\sqrt{\frac{r}{D}},
\end{equation}
then
\begin{equation}\label{key}
	\rho_1(x)=C_\alpha\frac{(z/2)^{\alpha+1}}{\Gamma(\alpha+1)}\frac{K_{\alpha+1}(z)}{\sqrt{Dr}}+\frac{|x|}{2D}\left[\mathcal{H}_\alpha(z)-1\right],
\end{equation}
where the constant $ C_\alpha $ is given by \eref{eq:C_alpha} and the function $ \mathcal{H}_\alpha(z) $ is defined as
\begin{equation}\label{eq:H}
	\mathcal{H}_\alpha(z)=z\left[K_\alpha(z)\mathrm{L}_{\alpha-1}(z)+K_{\alpha-1}(z)\mathrm{L}_{\alpha}(z)\right].
\end{equation}
Here $ K_\nu(y) $ is a modified Bessel function of the second kind, while $ \mathrm{L}_\nu(y) $ is a modified Struve function \cite{Abr-Steg}. A similar procedure can be followed for $ \rho_2(x) $, whereby we arrive at
\begin{equation}\label{key}
	\rho_2(x)=\frac{1}{2v}\left[1-\mathcal{H}_\alpha(z)\right].
\end{equation}
The final expression for the stationary distribution is thus
\begin{equation}\label{eq:P_st_expression}
	P(x)=\frac{1}{\eta_\alpha}\left\{C_\alpha\frac{(z/2)^{\alpha+1}}{\Gamma(\alpha+1)}\frac{K_{\alpha+1}(z)}{\sqrt{Dr}}+\left(\frac{1}{2v}-\frac{|x|}{2D}\right)\left[1-\mathcal{H}_\alpha(z)\right]\right\},
\end{equation}
see \eref{eq:eta} for the definition of $ \eta_\alpha $. Figure \ref{fig:stat} displays a few cases of $ P(x) $ and the agreement with the corresponding numerical simulations. We observe that smaller values of the shape parameter yield narrower distributions, with more pronounced peaks around $ x=0 $. This reflects the fact that, if the value of the scale parameter is fixed, Gamma distributions with smaller $ \alpha $ on average lead to shorter displacement phases, and long diffusive excursions thus become less likely.

We point out that for half-integer values of $ \alpha $, the stationary distribution can be written in terms of elementary functions, see \ref{app:P_st_explicit}. In particular, for $ \alpha=\case 12 $, viz., in the case of resetting at constant rate $ r $, equation \eref{eq:P_st_expression} reduces to
\begin{equation}\label{eq:P_st_poisson}
	P(x)=\frac 12\sqrt{\frac{r}{D}}\rme^{-|x|\sqrt{\frac{r}{D}}},
\end{equation}
which is independent of the return velocity $ v $ and equivalent to the result obtained in the case of instantaneous returns. However, we will show in the next section that in general the stationary distribution is dependent on $ v $.

\section{Dependence of the stationary distribution on the return velocity}\label{s:Pst_vdep}
The results of the previous section show, as it has already been proved in the literature \cite{PalKusReu-2019PRE,PalKusReu-2019,BodSok-2020-BrResI,RAD-2021}, that in the case of Poissonian resetting the stationary distribution is completely unaffected by the value of the return velocity, meaning that the return phase does not play any role in the definition of the steady state. Interestingly, this observation also applies to the finite-$ t $ behaviour of the PDF \cite{RAD-2021,PalKusReu-2019PRE}. However, when the waiting times for the resetting events are Gamma-distributed, this is true only for a specific value of the shape parameter. In particular, $ P(x) $ becomes $ v $-independent only for exponentially-distributed waiting times, suggesting that the requirement of a constant resetting rate represents a crucial condition for the independence of the steady state.

\begin{figure*}[h!]
	\centering
	\begin{tabular}{c @{\quad} c }
		\includegraphics[width=.45\linewidth]{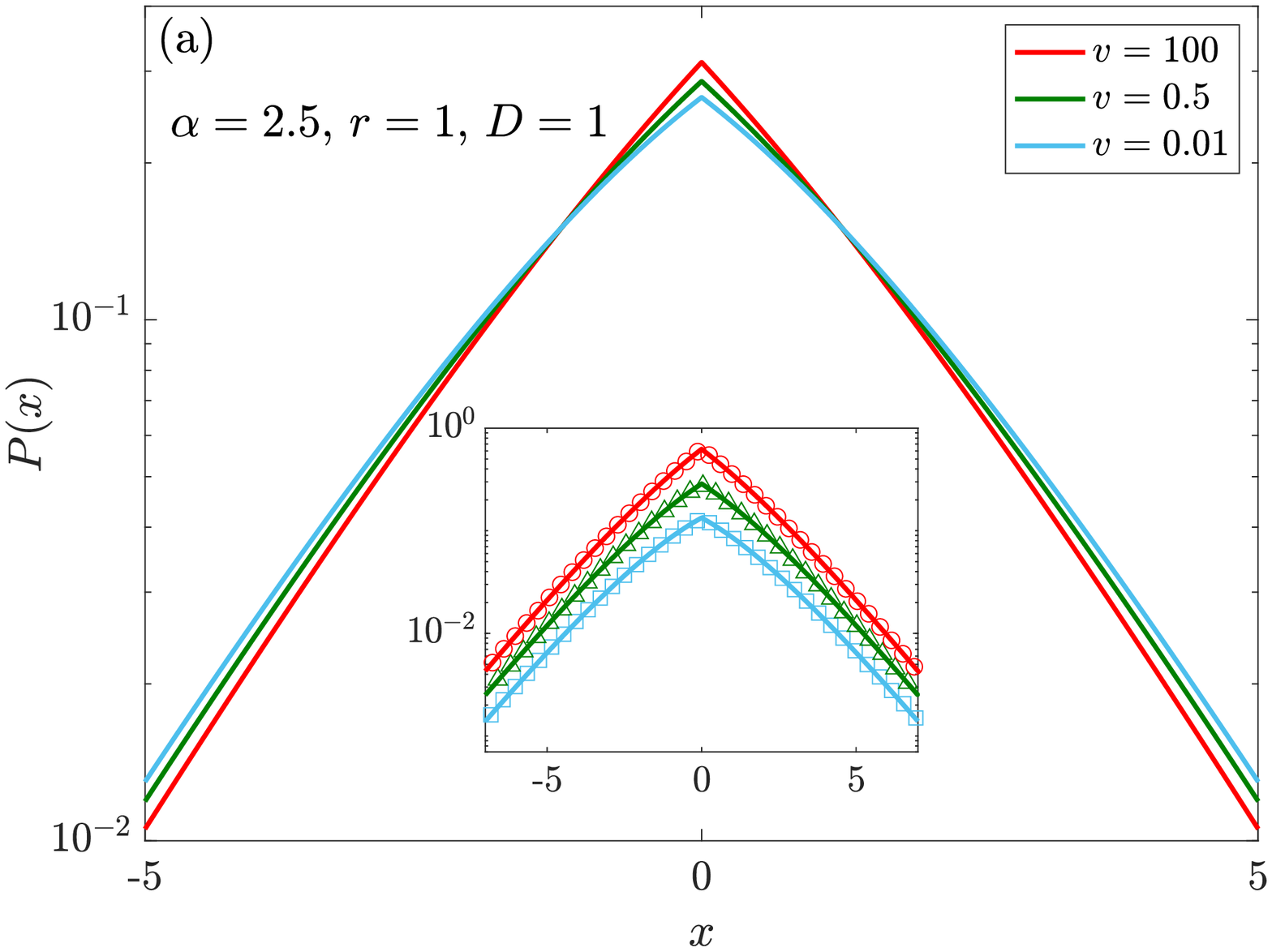} &
		\includegraphics[width=.45\linewidth]{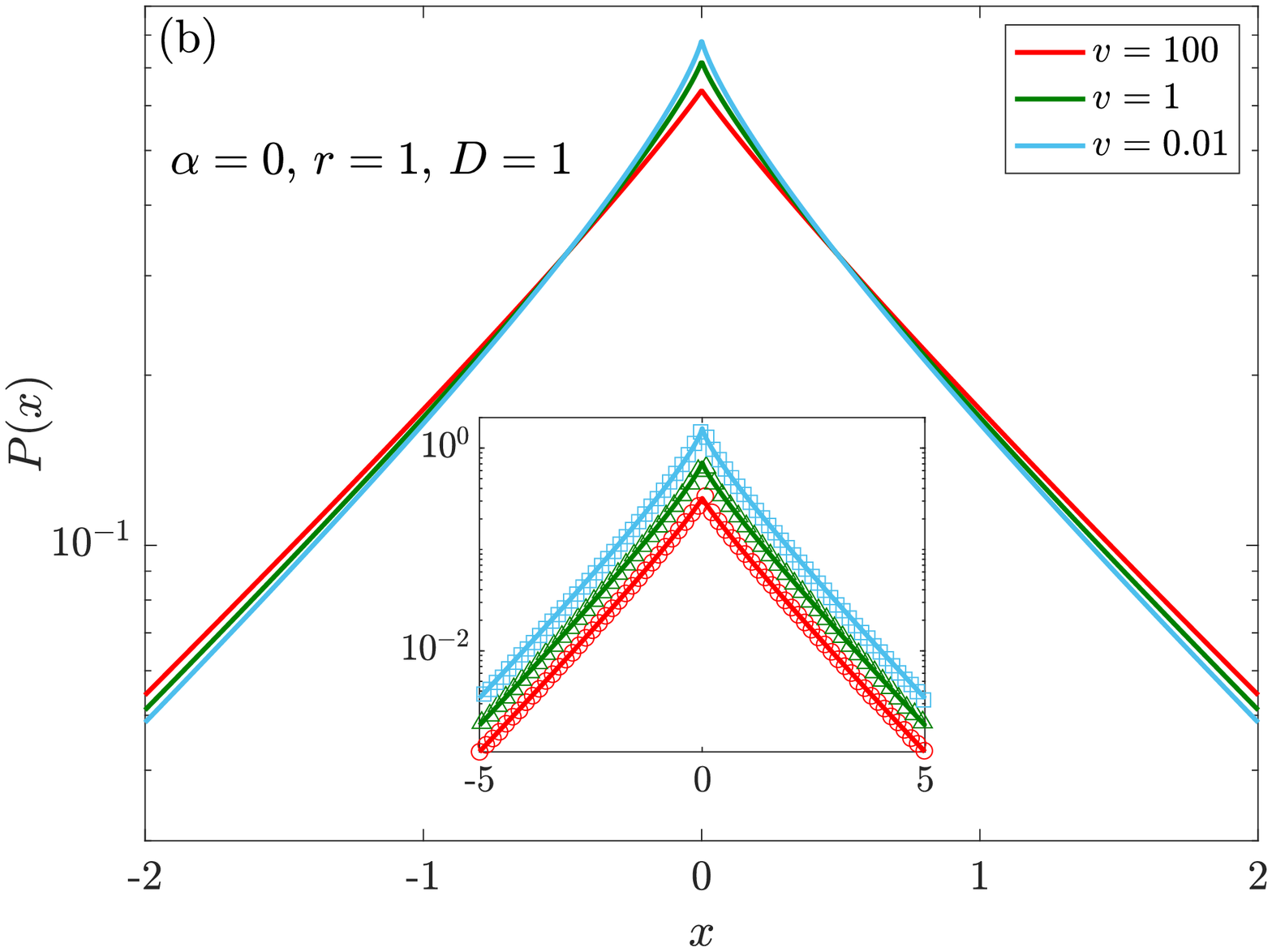} 
	\end{tabular}
	\includegraphics[width=.45\linewidth]{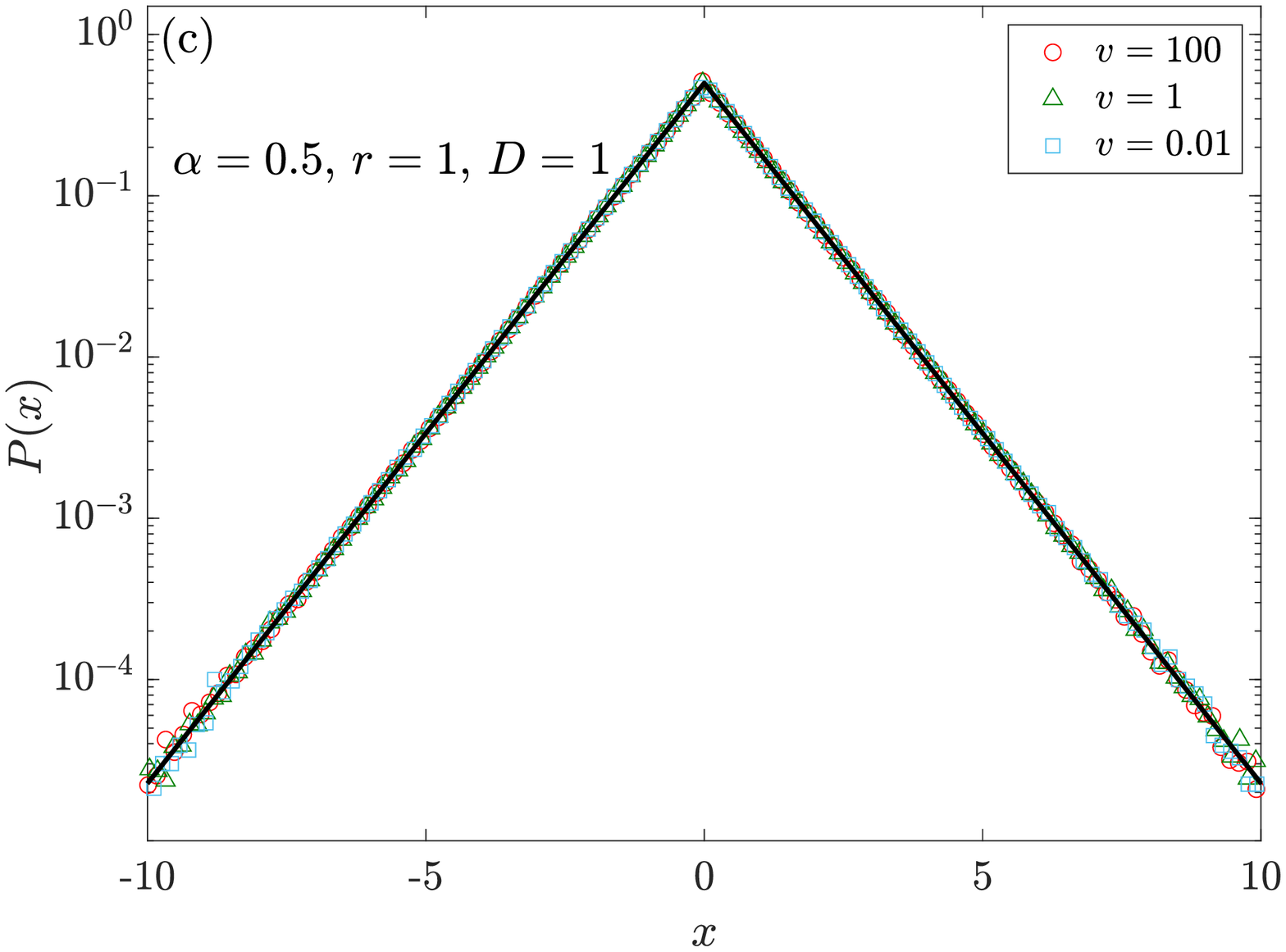}
	\caption{Examples of stationary distributions for different values of the return velocity $ v $ and shape parameter $ \alpha $. Panel (a): examples of $ P(x) $ for $ \alpha=2.5 $, with $ v=100 $ (red), $ v=0.5 $ (green) and $ v=0.01 $ (light blue). The agreement between numerical data and the theoretical curves of \eref{eq:P_st_expression} is shown in the inset. All data sets are obtained by simulating $ 10^7 $ walks with time step $ \rmd t=0.01 $ up to time $ t=10 $ ($ v=100 $, red circles and $ v=0.5 $, green triangles) or $ t=100 $ ($ v=0.01 $, light blue circles). Note that the data relative to $ v=100 $ and $ v=0.01 $ have been multiplied by  $2$ and $0.5$, respectively, to avoid overlap. Panel (b): the same as panel (a), but with $ \alpha=0 $ and $ v=100 $, $1$, $0.01$. In this case, the data in the inset are shifted by a factor $ 2 $ for $ v=0.01 $, and $ 0.5 $ for $ v=100 $, again with total evolution time $ t=10 $ ($ v=100 $, red circles and $ v=1 $, green triangles) and $ t=100 $ ($ v=0.01 $, light blue circles). Panel (c): stationary distribution for $ \alpha=0.5 $. In this case there is no dependence on the return velocity, indeed all data sets collapse on the same curve (solid black line), given by \eref{eq:P_st_poisson}. Here for any value of $ v $ the evolution time is $ t=10 $.}
	\label{fig:P_vdep}
\end{figure*}

Indeed, let us consider \eref{eq:P_st_expression} and evaluate $ P(x) $ around $ x=0 $. The behaviour of the modified Bessel function for $ \nu>0 $ is given by \cite{Abr-Steg}
\begin{equation}\label{eq:K_asymp_0}
	K_{\nu}(z)\sim\frac{\Gamma\left(\nu\right)}{2}\left(\frac 2z\right)^{\nu},\quad z\to0,
\end{equation}
and we recall that $ K_{\nu}(z)=K_{-\nu}(z) $. By using the definition of $ \mathrm{L}_{\nu}(y) $, see \cite{Abr-Steg}, it follows that as $ z\to 0 $
\begin{equation}\label{eq:H_asymp_0}
	\mathcal{H}_{\alpha}(z)\sim
	\cases{
	\frac{\Gamma(-\alpha)}{\sqrt{\pi}}\frac{\left( z/2\right)^{1+2\alpha}}{\Gamma\left(\alpha+\frac32\right)}&for $-\frac 12<\alpha<0$\\
	-\frac{2}{\pi}z\log z &for $\alpha=0$\\
	\frac{\Gamma(\alpha)}{\Gamma(\alpha+\frac12)}\frac{z}{\sqrt{\pi}}&for $\alpha>0$.
}
\end{equation}
It is then easy to see that $ P(x) $ attains a finite value at $ x=0 $, namely
\begin{equation}\label{key}
	P(0)=\frac 1{2C_\alpha}\sqrt{\frac rD}\cdot\frac{\sqrt{rD}+C_\alpha v}{\sqrt{rD}+B_\alpha v},
\end{equation}
where the constant $ B_\alpha $ is
\begin{equation}\label{key}
	B_\alpha=\frac{\alpha+1/2}{C_\alpha}=\frac{\sqrt{\pi}}{2}\frac{\Gamma\left(\alpha+\frac 32\right)}{\Gamma(\alpha+1)},
\end{equation}
and $ C_\alpha $ is given by \eref{eq:C_alpha}. For any $ v>0 $, the value of $ P(0) $ depends on $ v $ unless the two coefficients $ B_\alpha $ and $ C_\alpha $ are equal. This condition is equivalent to
\begin{equation}\label{key}
	\frac{\Gamma\left(\alpha+1\right)}{\Gamma(\alpha+\frac 12)}=\frac 12\sqrt{\pi(\alpha+1/2)},
\end{equation}
and one can verify that the only real solution to this equation is $ \alpha=\case 12 $. Therefore, taking $ \alpha=\case 12 $ is a necessary condition to obtain a steady state completely independent of the return velocity. In a similar fashion, we can also compute the behaviour of the stationary distribution for large values of $ |x| $. The asymptotic expansion of the modified Bessel function is \cite{Abr-Steg}
\begin{equation}\label{eq:K_asymp_inf}
	K_\nu(z)\sim\sqrt{\frac \pi{2z}}\rme^{-z}\left[1+\frac{4\nu^2-1}{8z}+o\left(\frac 1z\right)\right],
\end{equation}
and by using the expansion of $ \mathrm{L}_\nu(z) $, we obtain that for large $ z $
\begin{equation}\label{eq:H_asympt_inf}
	\mathcal{H}_\alpha(z)\sim 1-\left(\frac z2\right)^{\alpha-\frac 12}\frac{\rme^{-z}}{\Gamma\left(\alpha+\frac12\right)}\left[1+\frac{\mathcal{A}_\alpha}{z}+o\left(\frac 1z\right)\right],
\end{equation}
where the coefficient $ \mathcal{A}_\alpha $ is defined by
\begin{equation}
	\mathcal{A}_\alpha= \frac12\alpha^2+\alpha-\frac 58.
\end{equation}
It follows that the behaviour of the stationary distribution for large $ |x| $ reads
\begin{equation}\label{key}
	P(x)\sim \sqrt{\frac rD}\frac {(z/2)^{\alpha-\frac 12}}{\Gamma\left(\alpha+\frac 12\right)}\frac{\rme^{-z}}{2C_\alpha}\cdot\frac{\sqrt{rD}+v}{\sqrt{rD}+B_\alpha v},
\end{equation}
which is dependent on $ v $ unless one takes values of $ \alpha $ such that
\begin{equation}\label{key}
	\frac 2{\sqrt{\pi}}\frac{\Gamma(\alpha+1)}{\Gamma\left(\alpha+\frac 12\right)}=\alpha+\frac 12.
\end{equation}
This condition is once again satisfied only for $ \alpha=\case 12 $. We observe that this value of the shape parameter represents a crossover between two different regimes of the PDF. Indeed, if we consider a fixed $ -\case 12<\alpha<\case 12 $, it is possible to verify that for higher return velocities $ P(0) $ decreases, while the tails of the PDF attain higher values; contrarily, for fixed $ \alpha>\case 12 $ an increasing $ v $ yields more pronounced peaks around $ x=0 $ and less important tails, see figure \ref{fig:P_vdep}. As we have already mentioned, the value $ \alpha=\case 12 $ also represents a crossover for the behaviour of the rate $ r(t) $: as shown in figure \ref{fig:rates}, the rate is decreasing for $ \alpha<\case 12 $ and increasing for $ \alpha>\case 12 $. Hence the only case where the stationary distribution is independent of $ v $ is for constant resetting rates, namely for Poissonian resetting. Note that this condition is not only necessary, but also sufficient; we can thus conclude that when the return motion is performed at constant velocity, the system reaches a $ v $-independent steady state only when the resetting events are described by Poissonian statistics. We point out, however, that in this paper we are only considering Gamma distributions which are characterized by a particular form of the rate $ r(t) $: we do not exclude that a $ v $-independent stationary distribution can be reached for more general choices of $ r(t) $. Furthermore, it is also worth observing that for different kinds of return motion one gets more general expressions of the return velocity, e.g., depending explicitly on the position $ x $ and the location of the resetting event $ x_0 $. In principle, it is possible that even for more general forms of the velocity with respect to the constant case, one obtains a $ v $-independent steady state. Indeed, it has been shown \cite{RAD-2021,BodSok-2020-BrResI} that for returns at constant acceleration $ a $, with $ v(x,x_0)=\sqrt{2a|x_0-x|} $, in the case of Poissonian resetting the steady state is completely unaffected by the return motion, being indeed described by the same stationary distribution observed for instantaneous returns. Generalizing this result to time-dependent resetting rates is however not straightforward and may represent a tempting challenge for future work.

\section{Mean first passage time}\label{s:MFPT}
In various contexts it is often of great importance to characterize the efficiency of a given search process. For this purpose, it is useful to consider the first passage properties. Consider for example an ensemble of one dimensional Brownian particles starting their motion at $ x=0 $, in search for a target located at $ b>0 $. The problem of finding the PDF of the position at time $ t $, conditioned to the fact the target has not been reached yet, has been widely considered and one finds the solution \cite{Red}
\begin{equation}\label{eq:q:displ}
	q(x,t;b)=\frac 1{\sqrt{4\pi Dt}}\left\{\exp\left[-\frac{x^2}{4Dt}\right]-\exp\left[-\frac{(2b-x)^2}{4Dt}\right]\right\}.
\end{equation}
The integral of this quantity is the total probability that a particle does not reach the target up to time $ t $, namely the survival probability
\begin{equation}\label{eq:Q_displ}
	Q(t,b)=\int_{-\infty}^{b}q(x,t;b)\rmd x=\mathrm{erf}\left(\frac b{\sqrt{4Dt}}\right),
\end{equation}
where $ \mathrm{erf}(z) $ denotes the error function:
\begin{equation}\label{key}
	\mathrm{erf}(z)=\frac2{\sqrt{\pi}}\int_{0}^{z}\rme^{-t^2}\rmd t.
\end{equation}
The difference between the fraction of particles who have survived up to time $ t $ and those who have survived up to time $ t+\rmd t $ corresponds to the fraction of particles that have reached the target for the first time between $ t $ and $ t+\rmd t $. The first passage time density is thus defined as
\begin{equation}\label{eq:f_displ}
	f(t,b)=-\frac{\partial Q(t,b)}{\partial t}=\frac{b}{\sqrt{4\pi Dt^3}}\exp\left(-\frac {b^2}{4Dt}\right),
\end{equation}
showing the characteristic $ t^{-\frac32} $ decay whereby one observes an infinite MFPT. However, it has been broadly discussed in the literature that the introduction of a resetting mechanism optimizes the efficiency of the process in such a way that the MFPT attains a finite value. Furthermore, different choices of the resetting protocol lead to different values of the MFPT \cite{PalReu-2017} and so it is interesting to evaluate the first passage properties of the system under different resetting mechanisms. It is also worth observing that the time cost to perform the return to the starting location inevitably increases the MFPT \cite{ZhoXuDen-2021,RAD-2021,PalKusReu-2020,BodSok-2020-BrResI}.

To study the first passage properties of the model considered in this paper, we adopt the approach of \cite{CheSok-2018}. We call \textit{successful} a subprocess during which the particle hits the target placed at $ b>0 $, and \textit{unsuccessful} a subprocess in which the resetting happens before the particle can reach the target. Note that since we are considering one dimensional systems, the target can only be reached during the displacement phase. Let us denote with $ \varpi(t,b) $ the probability density of hitting the target at time $ t $ after the start of the successful subprocess and $ \varphi(t,b) $ the probability density of the duration of an unsuccessful subprocess. Then the probability density of ending the $ n $-th unsuccessful subprocess at time $ t $ is
\begin{equation}\label{key}
	\varphi_n(t,b)=\cases{
	\varphi(t,b) & for $n=1$\\
	\int_{0}^{t}\varphi_{n-1}(t',b)\varphi(t-t',b)\rmd t' & for $n\geq2$.}
\end{equation}
The probability of hitting the target for the first time at time $ t $ is equal to the probability that the $ n $-th unsuccessful subprocess ends at time $ t'<t $, and the target is then reached in a time $ t-t' $. Hence the first passage density of the process is
\begin{equation}\label{key}
	F(t,b)=\varpi(t,b)+\sum_{n=1}^{\infty}\int_{0}^{t}\varpi\left(t-t',b\right)\varphi_n\left(t',b\right)\rmd t',
\end{equation}
which is written more conveniently in Laplace space as
\begin{equation}\label{key}
	\hat{F}(s,b)=\hat{\varpi}(s,b)\sum_{n=0}^{\infty}\hat{\varphi}^n(s,b)=\frac{\hat{\varpi}(s,b)}{1-\hat{\varphi}(s,b)}.
\end{equation}
The MFPT can then be obtained by evaluating the derivative of $ \hat{F}(s,b) $:
\begin{equation}\label{eq:mean_FPT}
	\langle T\rangle=-\left.\frac{\mathrm{d}\hat{F}(s,b)}{\mathrm{d} s}\right\vert_{s=0}=-\frac{\hat{\varpi}'(0,b)}{1-\hat{\varphi}(0,b)}-\frac{\hat{\varpi}(0,b)\hat{\varphi}'(0,b)}{\left[1-\hat{\varphi}(0,b)\right]^2}.
\end{equation}

We can now observe that the density $ \varpi(t,b) $ is equal to the first passage density of the displacement phase multiplied by the probability of not having been reset up to time $ t $. Thus the Laplace transform $ \hat{\varpi}(s,b) $ reads
\begin{eqnarray}
	\hat{\varpi}(s,b)&=&\int_{0}^{\infty}e^{-st}\Psi(t)f(t,b)\mathrm{d}t\\
	&=&\int_{0}^{\infty}\frac{\rmd t}{\sqrt{4\pi Dt^3}}\frac{\Gamma\left(\alpha+\frac 12,rt\right)}{\Gamma\left(\alpha+\frac 12\right)}\exp\left(-st-\frac{b^2}{4Dt}\right),\label{eq:varpi_s}
\end{eqnarray}
see \eref{eq:Gamma_surv} and \eref{eq:f_displ}. The derivative with respect to $ s $ is
\begin{equation}\label{key}
	-\hat{\varpi}'(s,b)=\int_{0}^{\infty}\frac{\rmd t}{\sqrt{4\pi Dt}}\frac{\Gamma\left(\alpha+\frac 12,rt\right)}{\Gamma\left(\alpha+\frac 12\right)}\exp\left(-st-\frac{b^2}{4Dt}\right),\label{eq:varpi_s_derivative}
\end{equation}
and by evaluating \eref{eq:varpi_s} and \eref{eq:varpi_s_derivative} at $ s=0 $, we can obtain the expressions of $ \hat{\varpi}(0,b) $ and $ \hat{\varpi}'(0,b) $. These integrals can be computed by following the same procedure we used for $ \rho_1(x) $ in section \ref{s:Pst_calc}, whereby we get
\begin{eqnarray}\label{key}
	\hat{\varpi}(0,b)&=&1-\mathcal{H}_\alpha(w)\\
	\hat{\varpi}'(0,b)&=&\frac{w^2}{2r}\left[1-\mathcal{H}_\alpha(w)\right]-\frac{2C_\alpha}{r}\left(\frac w2\right)^{\alpha+2}\frac{K_{\alpha+1}(w)}{\Gamma(\alpha+1)},
\end{eqnarray}
where the variable $ w $ is
\begin{equation}\label{key}
	w=b\sqrt{\frac rD}.
\end{equation}
The density of the duration of an unsuccessful subprocess can be computed as follows: up to the resetting event, the position of the particle always stays below $ b $ and is thus distributed according to $ q(x,t;b) $, given by \eref{eq:q:displ}. At the time of the resetting, which happens at a random moment $ \tau $ with distribution $ \psi(\tau) $, the particle occupies a random position $ x_0 $, from which the deterministic return motion towards the origin starts. The duration of the return phase is $ \theta(x_0)=|x_0|/v $, hence by averaging over all $ \tau $ and $ x_0 $ one gets
\begin{equation}\label{key}
	\varphi(t,b)=\int_{0}^{\infty}\mathrm{d}\tau\psi(\tau)\int_{-\infty}^{b}\mathrm{d}x_0\delta\left[t-\tau-|x_0|/v\right]q(x_0,\tau;b).
\end{equation}
Note that this definition has the same structure of the equation for $ \phi(t) $, see \eref{eq:phi}, except for the fact that here we are using the survival PDF $ q(x,t;b) $ and the integration domain in $ x_0 $ is thus limited to $ (-\infty,b) $. The Laplace transform of this quantity reads
\begin{equation}\label{eq:varphi_LT}
	\hat{\varphi}(s,b)=\int_{-\infty}^{b}\rmd x_0\rme^{-|x_0|s/v}\int_{0}^{\infty}\rmd\tau\rme^{-s\tau}\psi(\tau)q(x_0,\tau;b),
\end{equation}
which can be evaluated at $ s=0 $ as
\begin{eqnarray}\label{key}
	\hat{\varphi}(0,b)&=&\int_{0}^{\infty}\psi(\tau)Q(\tau,b)\rmd\tau\\
	&=&\mathcal{H}_\alpha(w),
\end{eqnarray}
see \eref{eq:app:I4} in \ref{app:Integrals} and note that $ Q(t,b) $ is defined by \eref{eq:Q_displ}. By evaluating the derivative of \eref{eq:varphi_LT} for $ s=0 $ we obtain
\begin{eqnarray}\label{key}
	-\hat{\varphi}'(0,b)&=&\frac 1v\int_{-\infty}^{b}\rmd x_0|x_0|q(x_0,\tau;b)\int_{0}^{\infty}\rmd\tau\psi(\tau)\nonumber\\
	&&+\int_{-\infty}^{b}\rmd x_0q(x_0,\tau;b)\int_{0}^{\infty}\rmd\tau\tau\psi(\tau),
\end{eqnarray}
and these integrals can be computed with similar procedures of the previous ones, yielding
\begin{eqnarray}\label{key}
	-\hat{\varphi}'(0,b)&=&\frac{\alpha+1/2}{r}\mathcal{H}_{\alpha+1}(w)+\frac bv\bigg[1+\mathcal{H}_\alpha(w)-2\mathcal{H}_{\alpha}(2w)\nonumber\\
	&&+\frac{C_\alpha}{w}-\frac{2C_\alpha w^\alpha}{\Gamma(\alpha+1)} K_{\alpha+1}(2w)\bigg].
\end{eqnarray}
By using \eref{eq:mean_FPT} and rearranging terms, one finds that the MFPT can then be written as
\begin{equation}\label{eq:MFPT_sum}
	\langle T\rangle = \langle T_0\rangle+\frac{\langle x_b\rangle}{v},
\end{equation}
where $ \langle T_0\rangle $ is the MFPT in the case of instantaneous returns
\begin{eqnarray}\label{eq:MFPT_T0}
	\langle T_0\rangle &=&\frac{1}{1-\mathcal{H}_\alpha(w)}\bigg\{\frac{2C_\alpha}{r}\left(\frac w2\right)^{\alpha+2}\frac{K_{\alpha+1}(w)}{\Gamma(\alpha+1)}\nonumber\\
	&&+\frac{\alpha+1/2}{r}\mathcal{H}_{\alpha+1}(w)-\frac{w^2}{2r}\left[1-\mathcal{H}_\alpha(w)\right]\bigg\},
\end{eqnarray}
while the second term is the contribution of the return motion, written in terms of the average ballistic distance:
\begin{eqnarray}\label{eq:MFPT_xb}
	\langle x_b\rangle &=&\frac{b}{1-\mathcal{H}_\alpha(w)}\bigg\{\frac{C_\alpha}{w}-\frac{2C_\alpha w^\alpha}{\Gamma(\alpha+1)}K_{\alpha+1}(2w)\nonumber\\
	&&+1+\mathcal{H}_\alpha(w)-2\mathcal{H}_\alpha(2w)\bigg\}.
\end{eqnarray}
It is interesting to study the behaviour of $ \langle T\rangle $ for small and large values of $ w $. By using \eref{eq:K_asymp_0} and \eref{eq:H_asymp_0} we find that for $ w\approx 0 $
\begin{eqnarray}
	\langle T_0\rangle &\sim& \frac{C_\alpha w}{r}\\
	\langle x_b\rangle&\sim&b,
\end{eqnarray}
hence one obtains
\begin{equation}\label{key}
	\langle T\rangle\sim b\left(\frac{C_\alpha}{\sqrt{Dr}}+\frac 1v\right).
\end{equation}
Note that since $ w=b\sqrt{r/D} $, the limit $ w\to0 $ may be achieved with different limits of the system parameters and corresponds to different behaviours of the MFPT. For example, if we consider small values of $ r $ and fix $ D $ and $ b $, then the leading term is the contribution of the diffusive phase $ \langle T_0\rangle $, and the MFPT attains large values; on the other hand, if we take $ D\to\infty $ and fix the other parameters, then the leading term is the contribution of the return phase, and the MFPT attains the finite value $ b/v $ independently of $ r $. The same observation is valid in the opposite limit $ w\to \infty $, which may be checked by using the asymptotic expansions \eref{eq:K_asymp_inf} and \eref{eq:H_asympt_inf}. In this case we find
\begin{eqnarray}
	\langle T_0\rangle &\sim& \frac{\Gamma\left(\alpha+\frac32\right)}{(w/2)^{\alpha-\frac12}}\frac{\rme^w}{r}\\
	\langle x_b\rangle&\sim&b\cdot\frac{\Gamma\left(\alpha+1\right)}{(w/2)^{\alpha+\frac12}}\frac{\rme^w}{\sqrt{\pi}},
\end{eqnarray}
and the resulting behaviour of the MFPT is thus
\begin{equation}\label{key}
	\langle T\rangle\sim \frac{\Gamma\left(\alpha+\frac12\right)}{(w/2)^{\alpha-\frac12}}\rme^w\left(\frac{\alpha+1/2}{r}+\frac{C_\alpha}{v}\sqrt{\frac Dr}\right),
\end{equation}
where it is clear that the first term on the right-hand side is the contribution of the displacement phase, while the second is the contribution of the return motion. For fixed $ D $ and $ b $, in the limit of large $ r $ the dominant term comes from the return phase and the MFPT diverges as
\begin{equation}\label{key}
	\langle T\rangle\approx \exp\left[\sqrt{r}-\left(\alpha+\frac12\right)\log\sqrt{r}\right];
\end{equation}
if instead we consider the limit $ D\to0 $, then the leading term is the contribution of the stochastic phase and the MFPT grows as
\begin{equation}\label{key}
	\langle T\rangle\approx\exp\left[\frac 1{\sqrt{D}}+\left(\alpha-\frac12\right)\log\sqrt{D}\right].
\end{equation}

\begin{figure*}[h!]
	\centering
	\includegraphics[width=.55\linewidth]{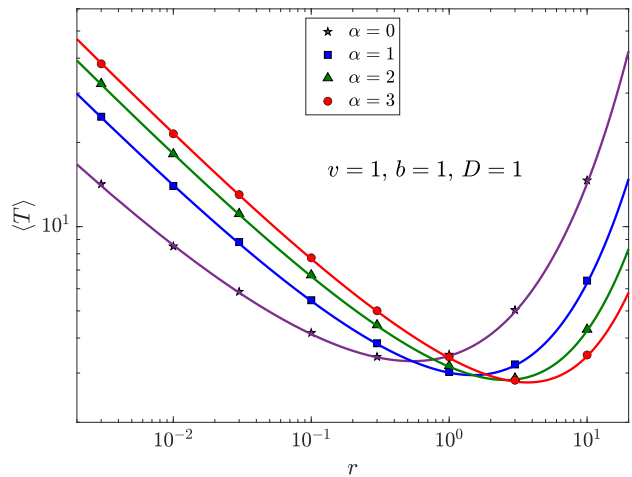}
	\caption{Mean first passage time as a function of the scale parameter $ r $ for various values of the shape $ \alpha $. All the numerical data sets are obtained by simulating $ 10^5 $ walks with small time step $ \rmd t=5\cdot10^{-4} $ and compared with the corresponding theoretical curves, see equations \eref{eq:MFPT_sum}, \eref{eq:MFPT_T0} and \eref{eq:MFPT_xb}, showing excellent agreement. In each case, the parameters $ b $, $D$ and $ v $ are all set to unity. As $ \alpha $ grows, the position of the minimum drifts towards bigger values, while its value becomes smaller.}
	\label{fig:MFPT_aInt}
\end{figure*}

In figure \ref{fig:MFPT_aInt} we present the results of our simulations on the MFPT. The theoretical curves and the data are shown as functions of the scale parameter $ r $ for a few values of the shape $ \alpha $, while the value of the target distance and the diffusion coefficient are fixed to unity. In every case, the agreement between data and theory is excellent. As we have already discussed, the curves diverge for both $ r\to0 $ and $r\to\infty$, and for each $ \alpha $ there exists an optimal value $ r^* $ that minimizes the MFPT. We observe that $r^*$ drifts towards higher values for bigger $ \alpha $, while the MFPT becomes smaller. This is consistent with some previous results in the literature which show that by fixing the mean duration of the displacement phase, the most effective resetting protocol corresponds to the limit $ \alpha\to\infty $, see \cite{EulMet-2016}. Since in our case the mean duration of the displacement phase is
\begin{equation}\label{key}
	\langle\tau\rangle=\frac{\alpha+1/2}{r},
\end{equation}
then the limit $ \alpha\to\infty $ with fixed $ \langle\tau\rangle $ implies that also $ r $ tends to infinity, hence one observes smaller values of the MFPT for higher $ \alpha $, in correspondence of bigger values of the optimal scale parameter $ r^* $. Furthermore, in the aforementioned limit the Gamma distribution becomes a sharp distribution centred around $ t=\langle\tau\rangle$. Indeed, by setting $ \alpha+\case12= r\langle\tau\rangle$, the density reads
\begin{equation}\label{key}
	\psi(t)=\frac{r\rme^{-rt}}{\Gamma\left(r\langle\tau\rangle\right)}\left(\frac t{\langle\tau\rangle}\right)^{r\langle\tau\rangle-1},
\end{equation}
thus in the large $ r $ limit we have
\begin{equation}\label{key}
	\psi(t)\sim\sqrt{\frac r{2\pi\langle\tau\rangle}}\exp\left[-r(t-\langle\tau\rangle)+\left(r\langle\tau\rangle-1\right)\log\left(\frac t{\langle\tau\rangle}\right)\right],
\end{equation}
which converges to zero as $ r\to\infty $ for $ t\neq\langle \tau\rangle $ and diverges for $ t=\langle\tau\rangle $. Hence the optimal resetting protocol corresponds to stochastic phases stopped after a fixed period $ \langle\tau\rangle $, as already discussed previously \cite{EulMet-2016,PalKunEva-2016,PalReu-2017,CheSok-2018}.

Similarly to what we have shown for the stationary distribution, we point out that the MFPT can be expressed in terms of elementary functions for half-integer values of the shape parameter, as we show in the following section, and use this simplification to investigate the dependence of the optimal scale $ r^* $ on the system parameters.

\section{Mean first passage time for half-integer values of the shape parameter and optimal scale}\label{s:MFPT_halfInt}
\begin{figure*}[h!]
	\centering
	\begin{tabular}{c @{\quad} c }
		\includegraphics[width=.45\linewidth]{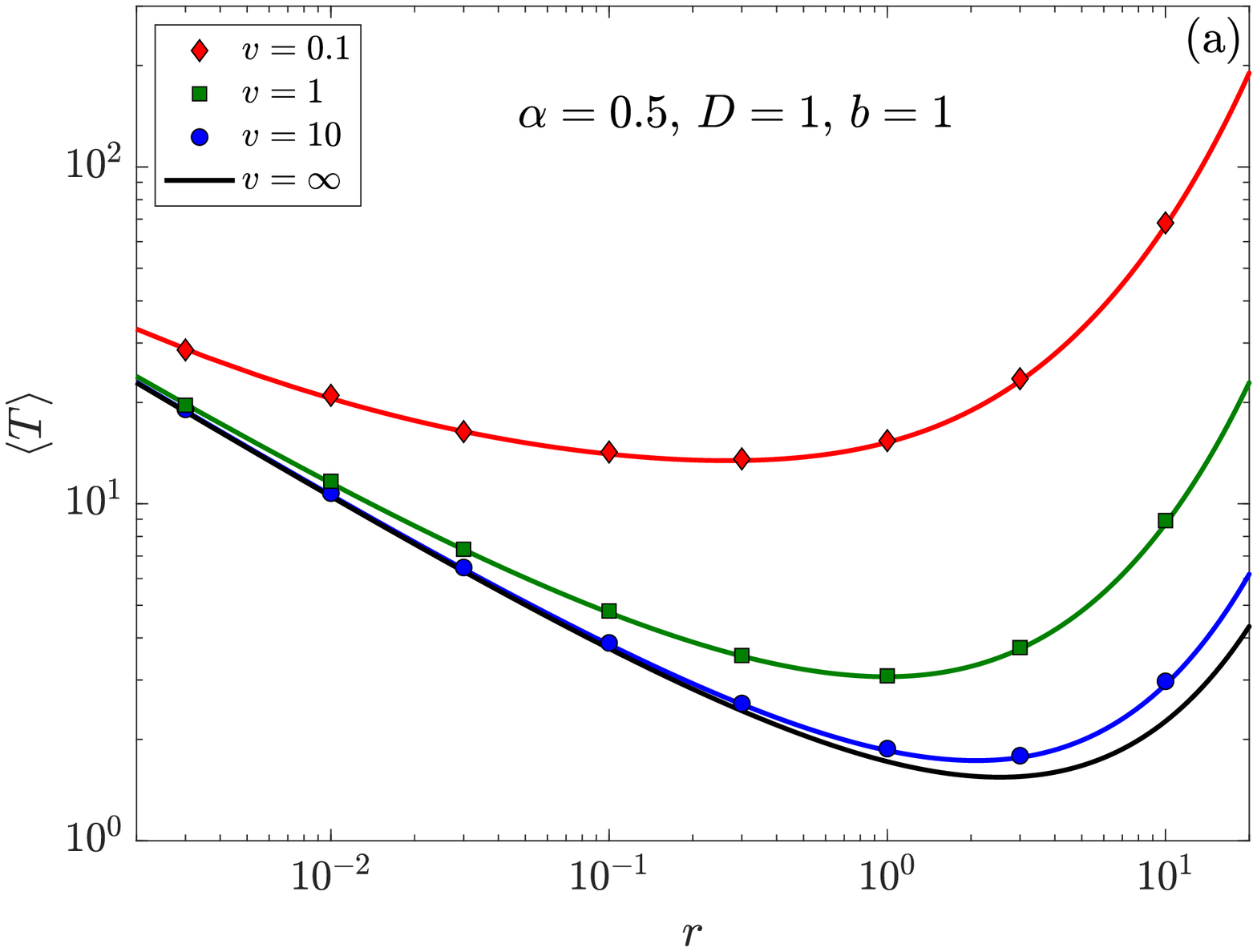} &
		\includegraphics[width=.45\linewidth]{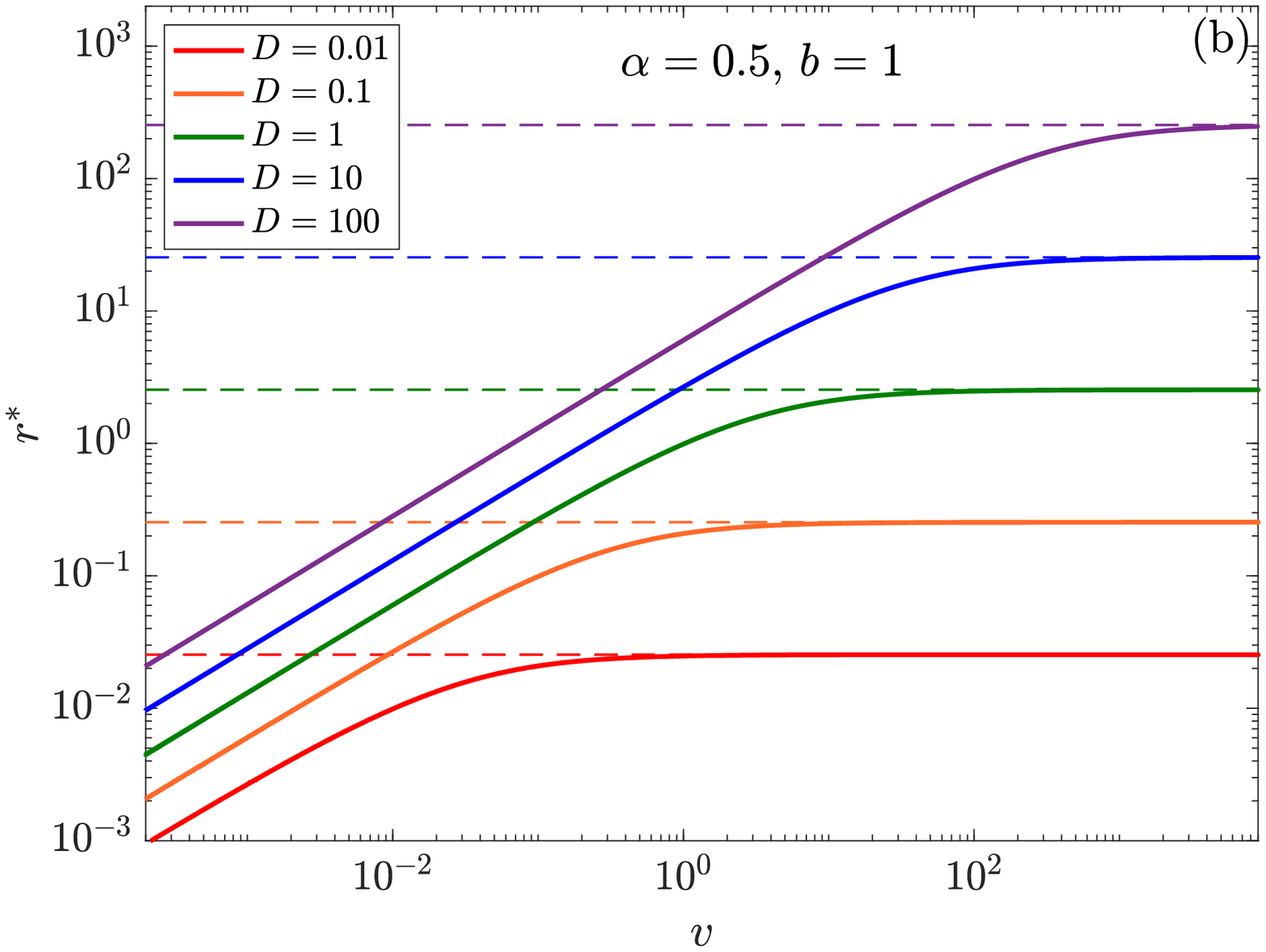} \\
		\includegraphics[width=.45\linewidth]{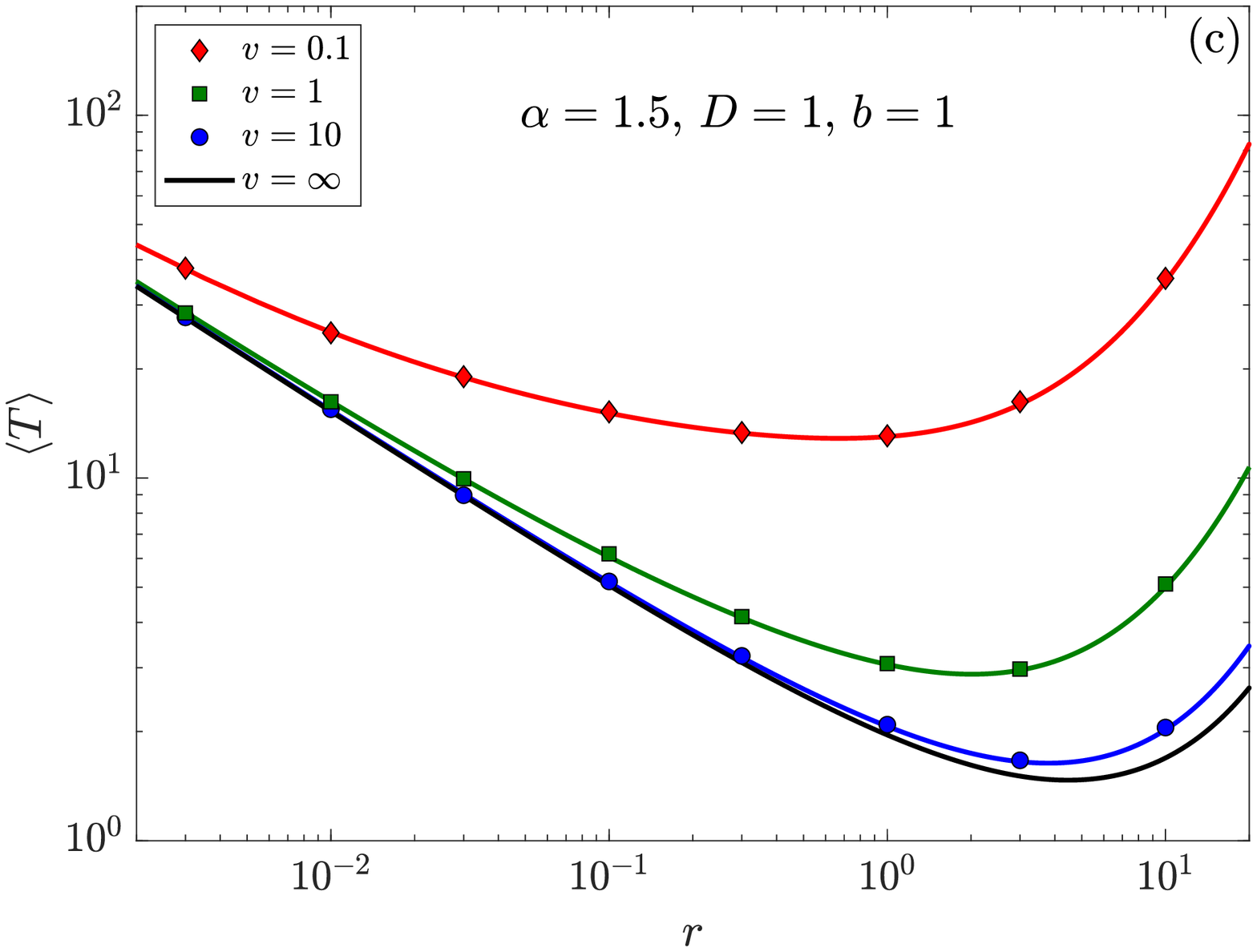} &
		\includegraphics[width=.45\linewidth]{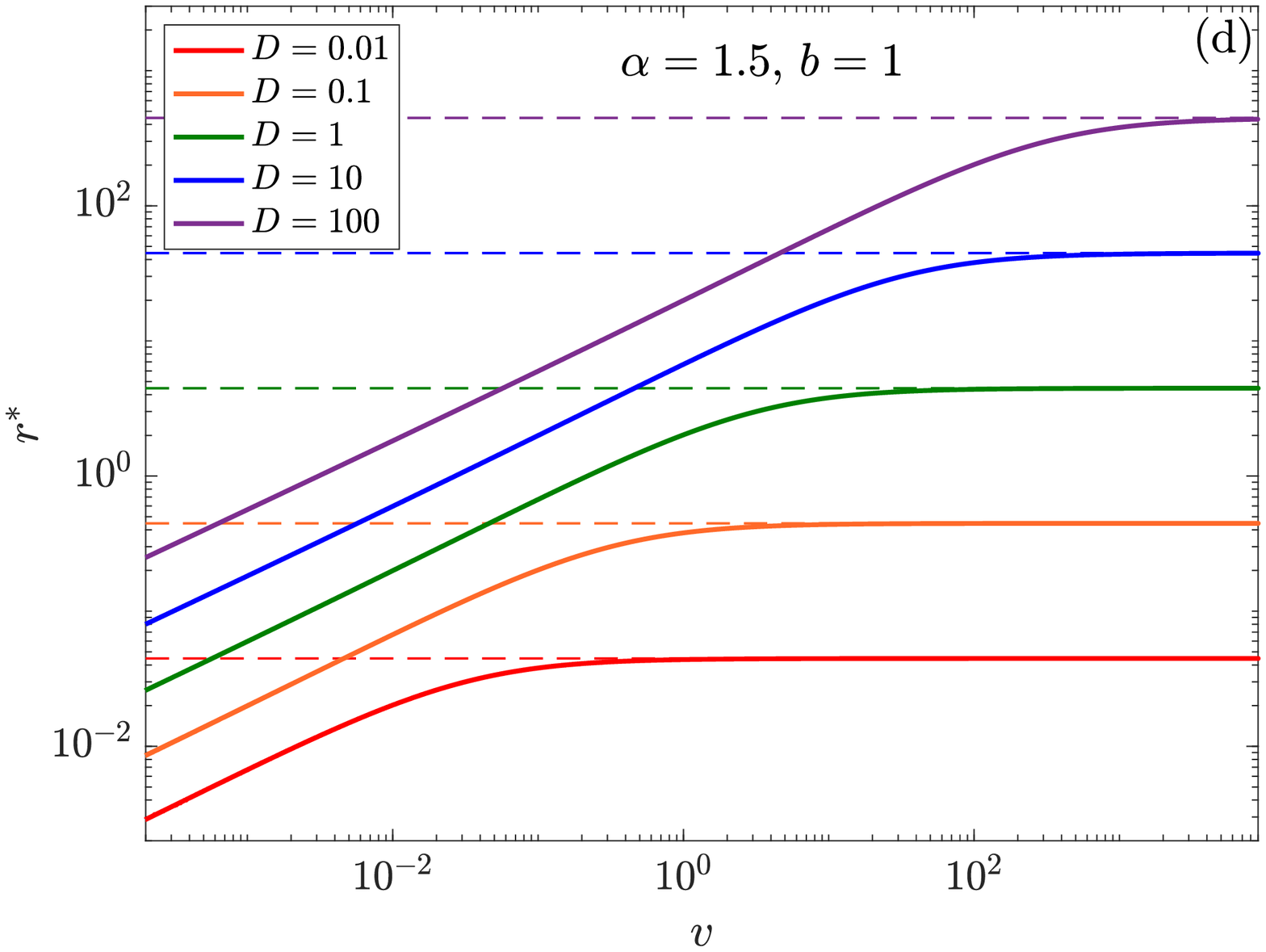}\\
		\includegraphics[width=.45\linewidth]{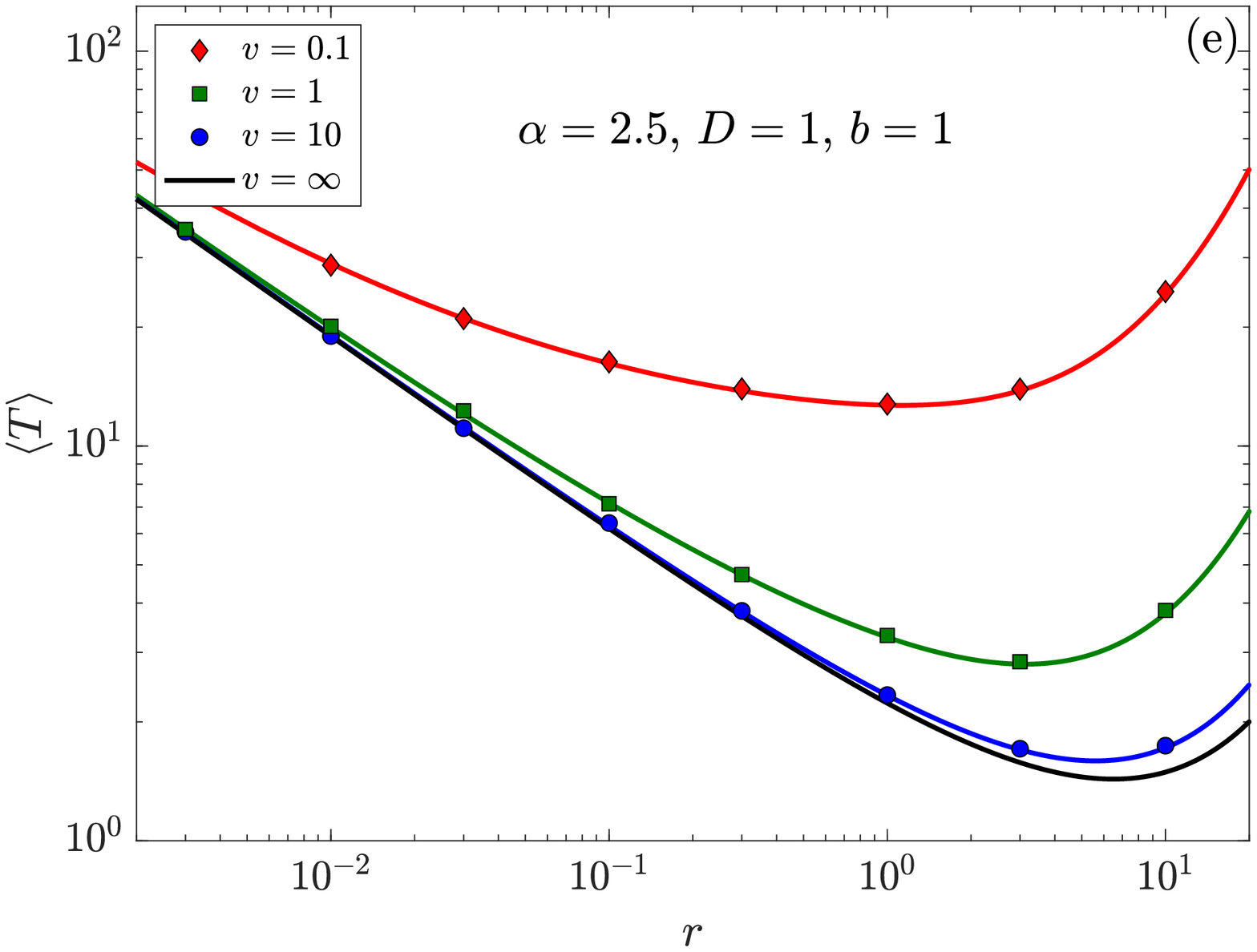} &
		\includegraphics[width=.45\linewidth]{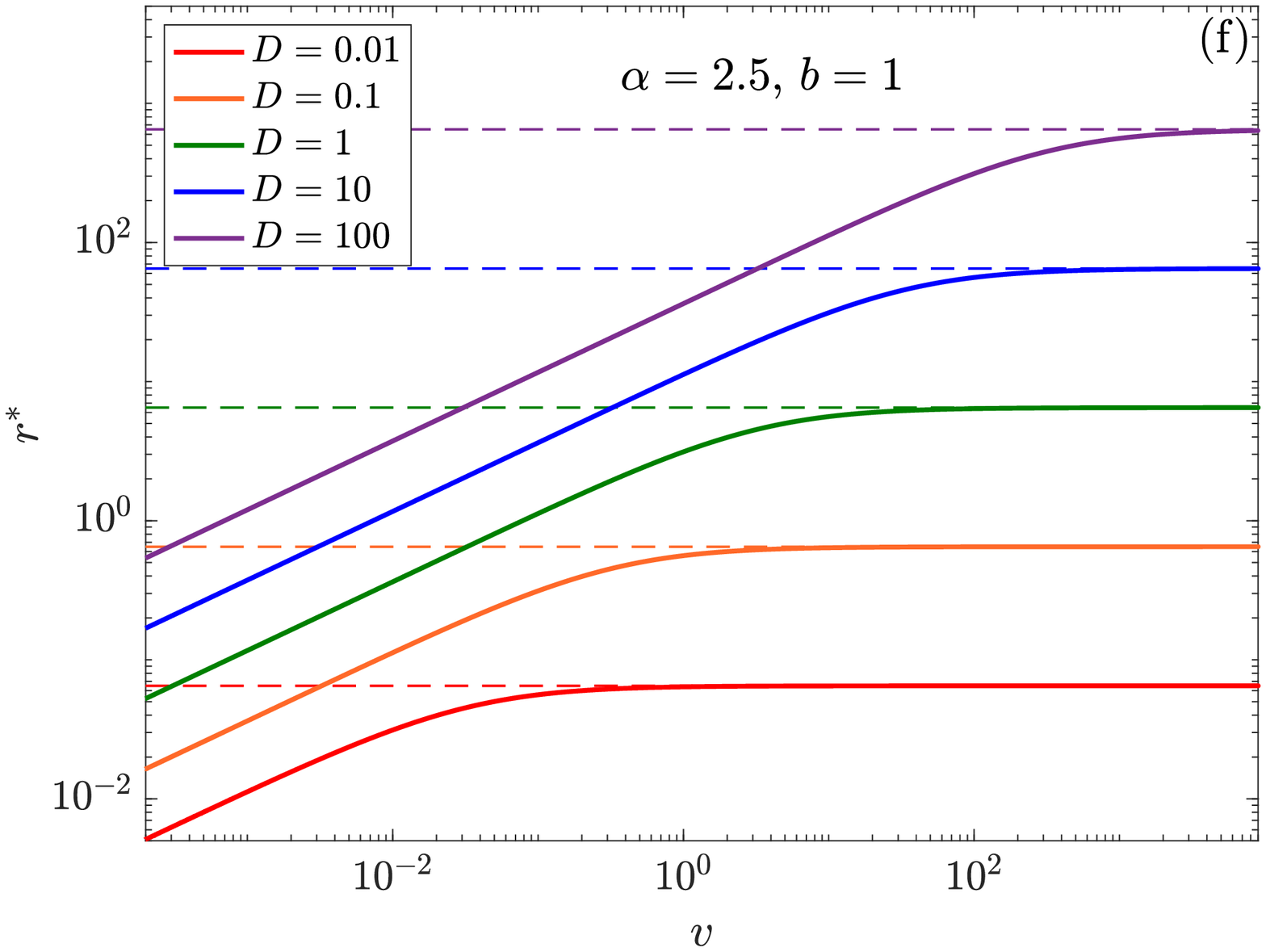}
	\end{tabular}
	\caption{Mean first passage time as a function of the scale parameter $ r $ (left panels) and dependence of the optimal scale parameter $ r^* $ on the velocity $ v $ (right panels), for half-integer values of the shape $ \alpha $. All the numerical data sets are obtained by simulating $ 10^5 $ walks with small time step $ \rmd t=5\cdot10^{-4} $. The solid black curves in the left panels represent the MFPT for instantaneous resetting, while the dashed horizontal lines in the right panels represent the values of $ r_0^* $ given by \eref{eq:opt_r0_numeric}.}
	\label{fig:MFPT_aSemInt}
\end{figure*}

As we show in \ref{app:P_st_explicit}, the functions appearing in the expression of $ \langle T\rangle $ can be simplified for $ \alpha=n+\case12 $. We report here only the main formulae, see the Appendix for more details. For $ m=-1,0,1,\dots, $ the modified Bessel function reads
\begin{equation}\label{key}
	K_{m+\frac 12}(z)=\sqrt{\frac{\pi}{2z}}\rme^{-z}R_m(z),
\end{equation}
where we define
\begin{equation}\label{key}
	R_m(z)=\cases{1&for $m=-1$\\
		\sum_{k=0}^{m}\frac{(m+k)!}{(m-k)!}\frac{(2z)^{-k}}{k!}&for $m\geq 0$.}
\end{equation}
Similarly, let us introduce
\begin{equation}\label{key}
	S_m(z)=\cases{0&for $m=-1$\\
		\frac 1{2^m}\sum_{k=0}^{m}\frac{(-1)^k(2k)!}{k!(m-k)!}z^{m-2k}&for $m\geq 0$.}
\end{equation}
Then for $ n=0,1,2,\dots, $ the function $ \mathcal{H}_{n+\frac12}(z) $ can be written as
\begin{eqnarray}\label{key}
	\mathcal{H}_{n+\frac 12}(z)&=1-\left[R_n(z)S_{n-1}(z)+R_{n-1}(z)S_{n}(z)\right]\rme^{-z}\\
	&=1-h_n(z)\rme^{-z}.
\end{eqnarray}
It follows that for $ \alpha=n+\case12 $ the quantities $ \langle T_0\rangle $ and $ \langle x_b\rangle $ can be expressed as
\begin{eqnarray}
	\langle T_0\rangle&=& \frac{n+1}{rh_n(w)}\left[\rme^w-h_{n+1}(w)-\frac{w^2h_n(w)}{2(n+1)}+\frac{w^{n+2}}{2^{n+1}}\frac{R_{n+1}(w)}{(n+1)!}\right]\label{eq:T_0_n}\\
	\langle x_b\rangle&=&\frac b{h_n(w)}\bigg\{\frac{\Gamma\left(n+\frac32\right)}{n!\Gamma\left(\frac32\right)}\frac{\rme^w}{w}-h_n(w)\nonumber\\
	&&+2\rme^{-w}\left[h_n(2w)-\frac{w^n}{n!}R_{n+1}(2w)\right]\bigg\}.\label{eq:x_b_n}
\end{eqnarray}
Panels (a), (c) and (e) in figure \ref{fig:MFPT_aSemInt} show the agreement of these results with our numerical simulations, which is excellent for all the considered values of $ \alpha $. For each shape parameter, we have considered different values of $ v $. Note that as $ v $ is increased, the curves approach the behaviour of the MFPT for instantaneous returns, represented in each panel by the solid black line. As we have already pointed out in the previous section, each curve has a single minimum $ \langle T^*\rangle $ in correspondence of the optimal value of the scale parameter, $ r^* $. As might be expected, for fixed $ \alpha $ the smallest of these minima, which we denote as $ \langle T_0^*\rangle $, is attained when the time cost for a return vanishes, i.e., in the limit $ v\to\infty $. In this case the second term at the right-hand side of \eref{eq:MFPT_sum} can be neglected and the optimal scale $ r_0^* $ is thus the solution of
	\begin{equation}\label{eq:T0_minimization}
		\frac{\partial\langle T_0\rangle}{\partial r}=0.
	\end{equation}
From \eref{eq:T_0_n} we observe that $ \langle T_0\rangle $ is always written as
\begin{equation}\label{key}
	\langle T_0\rangle = \frac 1rf(w),
\end{equation}
hence the condition \eref{eq:T0_minimization} can be recast only in terms of the variable $ w $:
\begin{equation}\label{eq:T0_minimization_w}
	\frac{\rmd}{\rmd w}\log f(w)=\frac 2w.
\end{equation}
We denote the positive solution of this equation with $ w_0^*$, and the optimal resetting rate $ r_0^* $ in the case of instantaneous returns is given by
	\begin{equation}\label{eq:opt_r0_numeric}
		r_0^*=D\left(\frac{w_0^*}{b}\right)^2,
	\end{equation}
which yields the minimum $ \langle T_0^*\rangle $ by plugging this value in the expression at the right-hand side of  \eref{eq:T_0_n}.

For finite $ v $ instead, the optimal $ r $ can be computed numerically. The dependence of $r^*$ on the velocity is shown in panel (b) for $ \alpha=\case12 $, panel (d) for $ \alpha=\case32 $ and panel (f) for $ \alpha=\case52 $. We note that as $ v $ is increased, $ r^* $ increases monotonically and converges to $ r_0^* $ in the limit $ v\to\infty $. The behaviour of $ r^* $ can be investigated by considering the equation
\begin{equation}\label{key}
	\frac{\partial \langle T\rangle}{\partial r}=\frac{\partial\langle T_0\rangle}{\partial w}\frac{\partial w}{\partial r}+\frac1v\frac{\partial\langle x_b\rangle}{\partial w}\frac{\partial w}{\partial r}=0,
\end{equation}
which can be written as
\begin{equation}\label{key}
	\frac{bv}D=\mathcal{F}_n(w),
\end{equation}
where the function $ \mathcal{F}_n(w) $ is defined by
\begin{equation}\label{key}
	\mathcal{F}_n(w)=-\frac{\partial \langle x_b\rangle}{\partial w}\cdot\left(\frac{\partial\langle T_0\rangle}{\partial w}\right)^{-1}.
\end{equation}
Here the subscript $ n $ denotes the dependence of both $ \langle T_0\rangle $ and $ \langle x_b\rangle $ on the shape parameter, see \eref{eq:T_0_n} and \eref{eq:x_b_n}. By fixing $ b $ and $ D $ the optimal value $ r^* $ can be computed in terms of the positive abscissa $ w^* $ for which the graph of the function $ \mathcal{F}_n(w) $ intersects the horizontal line $ y=bv/D $. The behaviour of $ r^* $ as a function of $ v $ can be thus deduced from the properties of $ \mathcal{F}_n(w) $, but to work with explicit expressions, in the following we consider a few particular cases of $ \alpha=n+\case12 $.

\subsection{Case $ \alpha=\case12 $}
Let us first consider the case of constant resetting rate, viz., Poissonian resetting. Then the formulae \eref{eq:T_0_n} and \eref{eq:x_b_n} simply yield
\begin{equation}\label{eq:MFPT_a0}
	\langle T\rangle=\frac1r\left(e^w-1\right)+\frac bv\left(\frac{2\sinh w}{w}-1\right),
\end{equation}
see indeed \cite{PalKusReu-2020,BodSok-2020-BrResI}. As explained previously, the optimal scale $ r_0^* $ in the case of instantaneous returns can be computed from \eref{eq:opt_r0_numeric} in terms of $ w_0^* $, namely the solution of equation \eref{eq:T0_minimization_w}, which in this case reads
\begin{equation}\label{key}
	\frac{\rme^w}{\rme^w-1}=\frac 2w.
\end{equation}
This equation is solved for $ w>0 $ by $ w_0^*\approx1.5936 $, and the smallest MFPT is thus
\begin{equation}\label{key}
	\langle T_0^*\rangle=\frac{1}{r_0^*}\left(\rme^{w_0^*}-1\right)\approx1.5441.
\end{equation}
In this case a deep analysis on the optimal $ r $ is proposed in \cite{PalKusReu-2020}. Here we just consider the equation for $ r^* $:
\begin{equation}\label{key}
	\frac{bv}D=\mathcal{F}_0(w),
\end{equation}
where the function $ \mathcal{F}_0(w) $ is 
\begin{equation}\label{key}
	\mathcal{F}_0(w)=\frac{2w\sinh w-2w^2\cosh w}{2-(2-w)\rme^w}.
\end{equation}
By equating the denominator to zero, it can be verified that $ \mathcal{F}_0(w) $ presents a divergence around $ w_0^* $. Moreover, it is possible to show that for $ w\in\left(0,w_0^*\right) $ the function is always positive and monotonic, furthermore as $ w\to0 $
\begin{equation}\label{key}
	\mathcal{F}_0(w)\sim\frac23w^3.
\end{equation}
Therefore the graph of the function always intersects the line $ y=bv/D $ at a single point, which approaches $ w_0^* $ as $ v $ becomes larger. From the monotonicity of $ \mathcal{F}_0(w) $ it follows that for smaller values of $ v $ one has smaller values of the optimal resetting rate, which explains the behaviour of the curves in panel (b).

\subsection{Case $ \alpha=\case32 $}
In this case we obtain
\begin{eqnarray}
	\langle T_0\rangle&=& \frac 1r\cdot\frac{4\rme^w-4-w}{2+w}\label{eq:T0_a3/2}\\
	\langle x_b\rangle&=&\frac bv\left(\frac{6\sinh w-2w\rme^{-w}}{2w+w^2}-1\right).
\end{eqnarray}
The optimal scale $ r_0^* $ for instantaneous returns is associated to $ w_0^* $, defined this time as the solution of
\begin{equation}\label{key}
	\rme^{-w}=\frac{8+2w-2w^2}{8+7w+w^2},
\end{equation}
which can be obtained from \eref{eq:T0_minimization_w} after rearranging terms. One can verify that it admits the positive solution $ w_0^*\approx2.1134 $, see panel (d), and the corresponding value of the smallest MFPT is thus
\begin{equation}\label{key}
	\langle T_0\rangle= \frac 1{r_0^*}\cdot\frac{4\rme^{w_0^*}-4-w_0^*}{2+w_0^*}\approx1.4691.
\end{equation}
Similarly to the case $ \alpha=\case12 $, the dependence on $ v $ of the optimal scale can be deduced from the equation
\begin{equation}\label{key}
	\frac{bv}{D}=\mathcal{F}_1(w),
\end{equation}
where $ \mathcal{F}_1(w) $ can be explicitly written as
\begin{equation}\label{key}
	\mathcal{F}_1(w)=\frac{\left(6w-3w^2\right)\sinh w-\left(6w^2+6w^3+w^4\right)\rme^{-w}}{8+7w+w^2-\left(8+2w-2w^2\right)\rme^w}.
\end{equation}
This function is positive and monotonic in $ \left(0,w_0^*\right) $, presents a divergence around $ w_0^* $, and as $ w\to0 $
\begin{equation}\label{key}
	\mathcal{F}_1(z)\sim w^2,
\end{equation}
see figure \ref{fig:plotF}, hence all the previous considerations also hold in this case. Once again for positive $ w $ the graph of the function has a single intersection with the line $ y=bv/D $, which yields the optimal value $ r^* $. These considerations lead to the monotonic behaviour of panel (d) in figure \ref{fig:MFPT_aSemInt}.

\subsection{Case $ \alpha=\case52 $}
\begin{figure*}[h!]
	\centering
	\begin{tabular}{c @{\quad} c }
		\includegraphics[width=.45\linewidth]{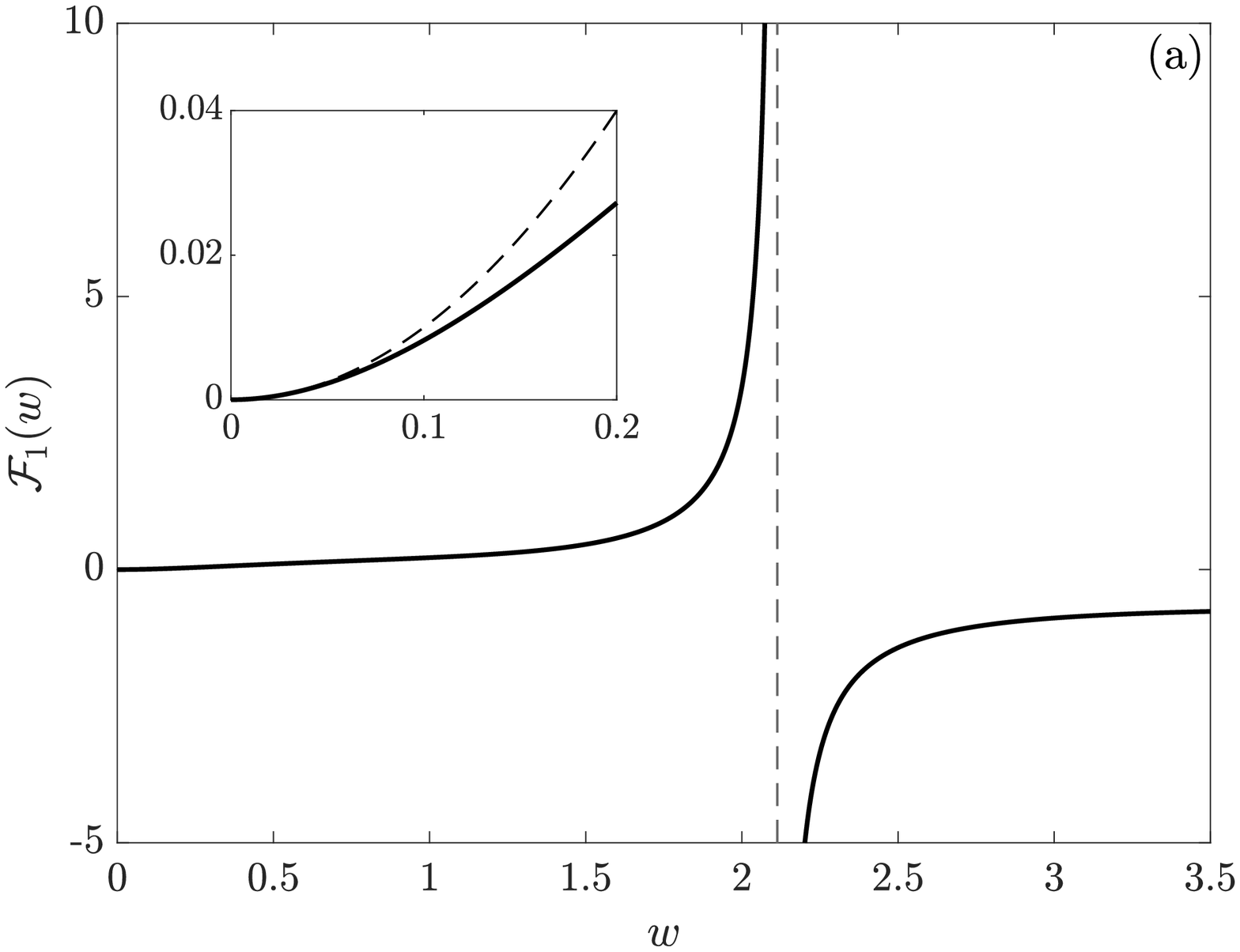} &
		\includegraphics[width=.45\linewidth]{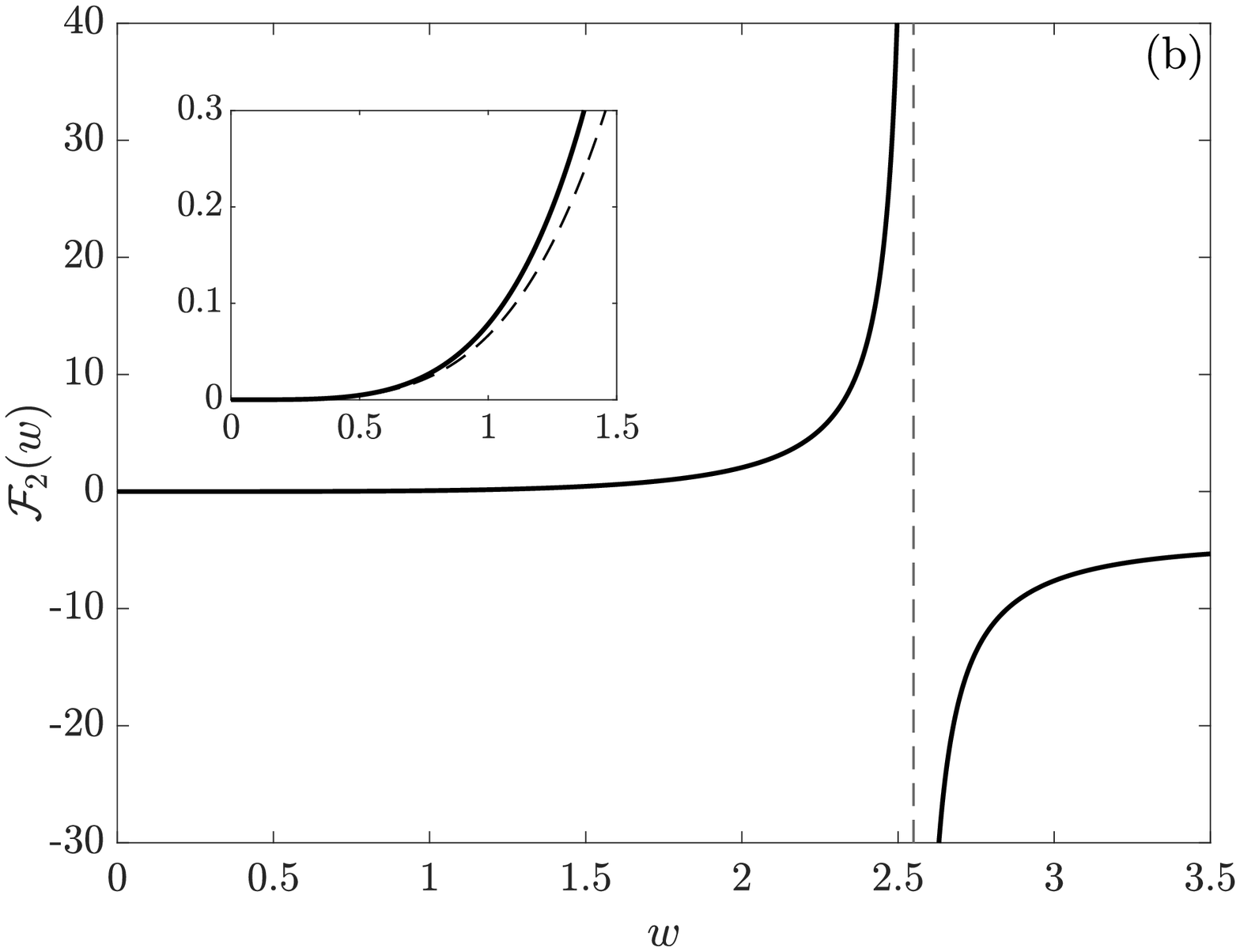}
	\end{tabular}
	\caption{Graphs of the functions $ \mathcal{F}_1(w) $ and $ \mathcal{F}_2(w) $ for positive values of the argument. The plots have been obtained by setting $ b $ and $ D $ to unity.  The insets show the functions for small values of $ w $ and their polynomial approximations, which are (a) $ \mathcal{F}_1(w)\sim w^2 $ and (b) $ \mathcal{F}_2(w)\sim w^4/15 $. The vertical dashed lines in both cases represent the value of $ w_0^* $.}
	\label{fig:plotF}
\end{figure*}

From \eref{eq:T_0_n} and \eref{eq:x_b_n} we get
\begin{eqnarray}
	\langle T_0\rangle&=& \frac 1r\cdot\frac{24\rme^w-24-9w-w^2}{8+5w+w^2}\label{eq:T0_a5/2}\\
	\langle x_b\rangle&=&\frac bv\left[\frac{30\sinh w-2w\rme^{-w}(7+2w)}{8w+5w^2+w^3}-1\right].
\end{eqnarray}
This time for $ v\to\infty $ the convergence of $ r^* $ is ruled by the positive solution of
\begin{equation}\label{key}
	w^3+w^2-7w-16+\left(16+18w+\frac{31}{4}w^2+\frac43w^3+\frac{1}{12}w^4\right)\rme^{-w}=0,
\end{equation}
which is $ w_0^*\approx2.5497 $. By computing the corresponding value of $ r_0^* $ and using \eref{eq:T0_a5/2}, we get
\begin{equation}\label{key}
	\langle T_0^*\rangle=\frac 1{r_0^*}\cdot\frac{24\rme^{w_0^*}-24-9w_0^*-(w_0^*)^2}{8+5w_0^*+(w_0^*)^2}\approx1.4329.
\end{equation}
The explanation of the dependence of $ r^* $ on the velocity follows the same line of the previous cases. This time the problem is related to the function
\begin{equation}\label{key}
	\mathcal{F}_2(w)=\frac{15wp_1(w)\rme^w-wp_2(w)\rme^{-w}}{2p_3(w)-24p_4(w)\rme^w},
\end{equation}
where $ p_i(w) $ are polynomials defined by
\begin{eqnarray}
	p_1(w)&=&8+2w-2w^2-w^3\\
	p_2(w)&=&120+270w+270w^2+145w^3+38w^4+4w^5\\
	p_3(w)&=&192+216w+93w^2+16w^3+w^4\\
	p_4(w)&=&16+7w-w^2-w^3.
\end{eqnarray}
The function is again monotonic and positive in $ \left(0,w_0^*\right) $, diverges around $ w_0^* $, and as $ w\to0 $
\begin{equation}\label{key}
	\mathcal{F}_2(w)\sim\frac{w^4}{15},
\end{equation}
see figure \ref{fig:plotF}, hence the behaviour of the curves in panel (f) of figure \ref{fig:MFPT_aSemInt} has the same origin discussed previously.

We remark that in agreement with the discussion of section \ref{s:MFPT}, for increasing values of $ \alpha $ we have computed increasing values of $ w_0^* $ and obtained decreasing values of $ \langle T_0^*\rangle $. In principle, the same analysis can be extended to any $ \alpha=n+\case 12 $, but more and more complex expressions should be expected as $ n $ grows.

\section{Conclusions}\label{s:Concl}
In this paper we have considered the problem of one dimensional diffusion with resetting and non-instantaneous returns. After the start of the process, a random waiting time is extracted from a Gamma distribution with shape parameter $ \alpha $ and scale parameter $ r $, after which the diffusive particle stops its motion and begins moving at constant velocity $ v $ towards the origin. Once the starting position is reached, the process starts anew. This is an example where the resetting mechanism requires a time cost, which has an effect to the statistical properties of the system.

In this research we have investigated the effects of the return motion on the stationary distribution, which shows an explicit dependence on $ v $. Such a dependence has been proved with both analytical results and numerical simulations. This represents a difference between previous results already studied in the literature, where it was shown that in the case of resetting performed at constant rate the steady state is unaffected by the return velocity. Indeed, the Gamma distribution possesses a nonconstant rate function, hence we can trace back this difference to the time dependence of the resetting rate. Even though the validity of this statement has been probed in this paper only for Gamma distributions, it would be interesting to consider the effects of the return motion on the stationary distribution for more general forms of the rate function.

We have also studied the search efficiency of the process by evaluating explicit expressions for the mean first passage time. For any value of $ \alpha $, the mean first passage time is minimized for a particular value $ r^* $ of the scale parameter, and both present a dependence on the return velocity. This is not surprising, since the time cost to complete the reset inevitably affects the search efficiency. Nevertheless, the system still shows a finite mean first passage time, contrarily to the corresponding reset-free process. It is worth observing that higher values of $ \alpha $ lead to smaller values of the mean first passage time, in correspondence however of larger values of the optimal scale parameter. This is consistent with the fact that, by fixing the mean waiting time to reset, the most efficient resetting protocol consists in fixed waiting periods: for the Gamma distribution, this is indeed achieved in the limit $ \alpha\to\infty $ and $ r\to\infty $ while keeping their ratio constant.

We think that this work can extend the knowledge on the matter of non-instantaneous resetting and supplement the literature on the problem of studying the features of nonconstant resetting rates, while also representing a bridge connecting these two areas of research.

\ack
The author acknowledges financial support from PRIN Research Project No. 2017S35EHN ``Regular and stochastic behaviour in dynamical systems'' of the Italian Ministry of Education, University and Research (MIUR).

\appendix
\section{Evaluation of some integrals in the main text}\label{app:Integrals}
In this Appendix we give the explicit solutions of some integrals we used to compute the main quantities presented in the main text. For $ \mathrm{Re}(p)>0 $ and $ \mathrm{Re}(y)>0 $ we have the following, see formula $6.561(4)$ in \cite{GraRyz}:
\begin{equation}\label{eq:app:I1}
	\int_{0}^{\infty}t^{\nu-1}\exp\left(-pt-\frac yt\right)\rmd t=2\left(\sqrt{\frac yp}\right)^{\nu}K_\nu\left(2\sqrt{py}\right),
\end{equation}
where $ K_\nu(z) $ is a modified Bessel function of the second kind.\\
For $ \mathrm{Re}(\nu)>\mathrm{Re}(\mu) $ and $ \mathrm{Re}(p+a)>0 $, it holds
\begin{eqnarray}
	\int_{0}^{\infty}t^{\nu-1}\rme^{-pt}K_\mu(at)\rmd t&=&\frac{\Gamma(\nu+\mu)\Gamma(\nu-\mu)}{\Gamma\left(\nu+\frac12\right)}\frac{\sqrt{\pi}(2a)^\mu}{(p+a)^{\nu+\mu}}\times\nonumber\\
	&&\hyp\left(\nu+\mu,\mu+\frac12;\nu+\frac12;\frac{p-a}{p+a}\right)\label{eq:app:I2},
\end{eqnarray}
see formula $6.621(3)$ in \cite{GraRyz}.\\
We move on to the following:
\begin{equation}\label{eq:app:I3}
	\int_{0}^{t}\frac{\rmd t'}{\sqrt{t'}}\exp\left(-\frac{y^2}{t'}\right)\rmd t=2\sqrt{t}\exp\left(-\frac{y^2}{t}\right)-2y\sqrt{\pi}\mathrm{erfc}\left(\frac{|y|}{\sqrt{t}}\right).
\end{equation}
This result can be easily proved by first considering the change of variable $ t'=(y/u)^2 $ and then integrating by parts the resulting integral.\\
Next we have, for $ \mathrm{Re}(\nu)>-\case 12 $ and $ y>0 $:
\begin{equation}\label{eq:app:I4}
	\int_{0}^{\infty}t^{\nu-\frac12}\rme^{-t}\mathrm{erfc}\left(\frac{y}{\sqrt{t}}\right)\rmd t=\Gamma\left(\nu+1/2\right)\left[1-\mathcal{H}_\nu(2y)\right],
\end{equation}
where $ \mathcal{H}_\nu(z) $ is defined by \eref{eq:H} in the main text. To prove formula \eref{eq:app:I4} we first use the definition of the complementary error function to write
\begin{equation}\label{key}
	\mathrm{erfc}(z)=1-\mathrm{erf}(z)=1-\frac2{\sqrt{\pi}}\int_{0}^{z}\rme^{-t^2}\rmd t.
\end{equation}
By plugging this expression in the integral on the lhs of \eref{eq:app:I4} one can write
\begin{equation}\label{key}
\fl	\int_{0}^{\infty}t^{\nu-\frac12}\rme^{-t}\mathrm{erfc}\left(\frac{y}{\sqrt{t}}\right)\rmd t=\Gamma(\nu+1/2)-\frac 2{\sqrt{\pi}}\int_{0}^{\infty}\rmd tt^{\nu-\frac12}\rme^{-t}\int_{0}^{y/\sqrt{t}}\rmd u\rme^{-u^2}.
\end{equation}
At this point we can use the change of variable $ u=wy/\sqrt{t} $ and perform the integral over $ t $ first, which is similar to the one given by \eref{eq:app:I1}, to obtain
\begin{equation}\label{key}
	\fl	\int_{0}^{\infty}t^{\nu-\frac12}\rme^{-t}\mathrm{erfc}\left(\frac{y}{\sqrt{t}}\right)\rmd t=\Gamma(\nu+1/2)-\frac {4y^{\nu+1}}{\sqrt{\pi}}\int_{0}^{1}w^\nu K_\nu(2yw)\rmd w.
\end{equation}
The integral in $ w $ is finally given by formula $ 6.561(4) $ in \cite{GraRyz}, and we arrive at the expression of \eref{eq:app:I4}.

\section{Stationary distribution for half-integer values of the shape parameter}\label{app:P_st_explicit}
We first consider the function
\begin{equation}
	\mathcal{H}_\alpha(z)=z\left[K_\alpha(z)\mathrm{L}_{\alpha-1}(z)+K_{\alpha-1}(z)\mathrm{L}_{\alpha}(z)\right],
\end{equation}
see \eref{eq:H} in the main text. Let us take $ n=-1,0,1,\dots, $ and define:
\begin{equation}\label{key}
	S_n(z)=\cases{0&for $n=-1$\\
	\frac 1{2^n}\sum_{k=0}^{n}\frac{(-1)^k(2k)!}{k!(n-k)!}z^{n-2k}&for $n\geq 0$.}
\end{equation}
For $ \alpha=n+\case 12 $, $ n=-1,0,1,\dots, $ the modified Struve function can be written as \cite{NIST}:
\begin{eqnarray}\label{key}
\mathrm{L}_{n+\frac 12}(z)&=I_{-n-\frac 12}(z)-\sqrt{\frac{2}{\pi z}}S_n(z)\\
&=I_{n+\frac 12}(z)+\frac 2\pi\sin\left(\frac\pi2+n\pi\right)K_{n+\frac 12}(z)-\sqrt{\frac{2}{\pi z}}S_n(z),
\end{eqnarray}
where $ I_\nu(y) $ and $ K_\nu(y) $ are modified Bessel functions of the first and second kind, respectively. We note that the function $ K_{n+\frac12}(z) $ can be expressed as follows \cite{NIST}:
\begin{equation}\label{key}
	K_{n+\frac 12}(z)=\sqrt{\frac{\pi}{2z}}\rme^{-z}R_n(z),
\end{equation}
where this time we define
\begin{equation}\label{key}
	R_n(z)=\cases{1&for $n=-1$\\
		\sum_{k=0}^{n}\frac{(n+k)!}{(n-k)!}\frac{(2z)^{-k}}{k!}&for $n\geq 0$.}
\end{equation}
Let us consider now $ \mathcal{H}_{n+\frac 12}(z) $ for $ n\geq 0 $. By using the relation \cite{Abr-Steg}
\begin{equation}\label{key}
	K_{\nu}(z)I_{\nu-1}+K_{\nu-1}I_{\nu}(z)=\frac 1z,
\end{equation}
we are able to write
\begin{eqnarray}\label{key}
	\mathcal{H}_{n+\frac 12}(z)&=1-\rme^{-z}\left[R_n(z)S_{n-1}(z)+R_{n-1}(z)S_{n}(z)\right]\\
	&=1-\rme^{-z}h_n(z).
\end{eqnarray}
Therefore for $ \alpha=n+\case 12 $, $ n\geq0 $, the expression of the stationary distribution, see \eref{eq:P_st_expression}, simplifies to
\begin{equation}\label{eq:app:P_simplified}
	P(x)=\frac{\rme^{-z}}{\eta_{\alpha}}\left[\left(\frac z2\right)^{n+1}\frac{R_{n+1}(z)}{n!\sqrt{Dr}}+\left(\frac 1{2v}-\frac{|x|}{2D}\right)h_n(z)\right],
\end{equation}
where $ z $ is defined as in the main text, $ z=|x|\sqrt{r/D} $. We can compute explicitly $ P(x) $ for the first few values of $ n $ by evaluating the corresponding values of $ R_n(z) $ and $ h_n(z) $. We have
\begin{eqnarray}\label{key}
	R_0(z)&=&1\\
	R_1(z)&=&1+\frac 1z\\
	R_2(z)&=&1+\frac 3z+\frac 3{z^2}\\
	R_3(z)&=&1+\frac 6z+\frac{15}{z^2}+\frac{15}{z^3},
\end{eqnarray}
and
\begin{eqnarray}\label{key}
	h_0(z)&=&1\\
	h_1(z)&=&1+\frac z2\\
	h_2(z)&=&1+\frac {5z}{8}+\frac {z^2}{8},
\end{eqnarray}
while the subprocess mean durations are
\begin{eqnarray}\label{key}
	\eta_{\frac 12}&=&\frac 1r+\frac 1v\sqrt{\frac Dr}\\
	\eta_{\frac 32}&=&\frac 2r+\frac 3{2v}\sqrt{\frac Dr}\\
	\eta_{\frac 52}&=&\frac 3r+\frac {15}{8v}\sqrt{\frac Dr}.
\end{eqnarray}
Therefore for $ \alpha=\case 12 $ the stationary distribution is
\begin{equation}\label{key}
	P(x)=\frac 12\sqrt{\frac rD}\rme^{-z},
\end{equation}
as previously observed in the literature. For $ \alpha=\case 32 $ we get 
\begin{equation}\label{key}
	P(x)=\frac 18\sqrt{\frac rD}\rme^{-z}\left[z+3+\frac{\sqrt{Dr}}{3\sqrt{Dr}+4v}(z-1)\right],
\end{equation}
which in the limit $ v\to\infty $ yields the stationary distribution in the case of instantaneous returns, which has been already proposed in \cite{EulMet-2016}. Finally for $ \alpha=\case 52 $ one obtains
\begin{equation}\label{key}
	\fl P(x)=\frac 1{16}\sqrt{\frac rD}\rme^{-z}\left[\frac{z^2}{3}+\frac{7z}{3}+5+\frac{5\sqrt{Dr}}{5\sqrt{Dr}+8v}\left(\frac{z^2}{5}+\frac z3-\frac {11}{15}\right)\right],
\end{equation}
and the expressions of the stationary distribution for each other value of $ n $ can be similarly evaluated.

\section{Simulation method}\label{app:simulation}
The analytical results in the main text for the PDF and the MFPT have been compared with numerical simulations. The evolution time is discretized with a small time step $ \Delta t $. The initial condition for the position is always $ x(0)=0 $. Upon the start of the displacement phase, a random number $ \tau $ is extracted from a Gamma distribution $\psi(\tau) $ with given shape and rate, representing the time of the first resetting event. The position then starts evolving following
\begin{equation}\label{app:eq:evol_disp}
	x(t+\Delta t)=x(t)+\eta\sqrt{2D\Delta t},
\end{equation}
where $ \eta $ is a random variable extracted from the standard normal distribution. When the evolution time reaches $ \tau $ the particle starts moving towards the origin according to
\begin{equation}\label{key}
	x(t+\Delta t)=x(t)-\mathrm{sgn}\left[x(\tau)\right]v\Delta t.
\end{equation}
As the origin is crossed, the process is restored to the initial condition $ x(0)=0 $, a new resetting time $ \tau $ is extracted and the previous steps are repeated up to the total observation time. Note that for the simulations on the MFPT it is not necessary to evolve the trajectory during the return phase: in this case one only needs the first passage time for each trajectory, hence it is sufficient to restore immediately the position to the initial condition and increase a counter for the first passage time by the cost $ \theta=|x(\tau)|/v$.

As a side note, instead of extracting the resetting time at the start of the displacement phase, it is also possible to start evolving according to \eref{app:eq:evol_disp} and checking after each time step whether the motion has to be switched to the return phase, according to the probability $ P=r(t)\Delta t $, where the rate function $ r(t) $ is defined by \eref{eq:Gamma_rate} in the main text. Both methods yield correct results, however we found the second method less efficient, mainly due to the repeated computations of the incomplete gamma function.

\section*{References}
\providecommand{\newblock}{}

%\bibliographystyle{iopart-num}
%\bibliography{/Users/mattia/Documents/Sorgenti_Tex/Biblio.bib}

\begin{thebibliography}{10}
	\expandafter\ifx\csname url\endcsname\relax
	\def\url#1{{\tt #1}}\fi
	\expandafter\ifx\csname urlprefix\endcsname\relax\def\urlprefix{URL }\fi
	\providecommand{\eprint}[2][]{\url{#2}}
	
	\bibitem{EvaMaj-2011}
	Evans M~R and Majumdar S~N 2011 {\em Phys. Rev. Lett.\/} {\bf 106} 160601
	
	\bibitem{DurHenPar-2014}
	Durang X, Henkel M and Park H 2014 {\em J. Phys. A: Math. Gen.\/} {\bf 47}
	045002
	
	\bibitem{EvaMaj-2014}
	Evans M~R and Majumdar S~N 2014 {\em J. Phys. A: Math. Theor.\/} {\bf 47}
	285001
	
	\bibitem{MajSabSch-2015}
	Majumdar S~N, Sabhapandit S and Schehr G 2015 {\em Phys. Rev. E\/} {\bf 91}
	052131
	
	\bibitem{NagGup-2016}
	Nagar A and Gupta S 2016 {\em Phys. Rev. E\/} {\bf 93} 060102(R)
	
	\bibitem{EvaMaj-2018Refractory}
	Evans M~R and Majumdar S~N 2018 {\em J. Phys. A: Math. Theor.\/} {\bf 52}
	01LT01
	
	\bibitem{BasKunPal-2019}
	Basu U, Kundu A and Pal A 2019 {\em Phys. Rev. E\/} {\bf 100} 032136
	
	\bibitem{MerBoyMaj-2020}
	Mercado-V\'{a}squez G, Boyer D, Majumdar S~N and Schehr G 2020 {\em J. Stat.
		Mech.\/} {\bf 113203}
	
	\bibitem{Bre-2021}
	Bressloff P~C 2021 {\em J. Phys. A: Math. Theor.\/} {\bf 54} 354001
	
	\bibitem{Gra-2021}
	Grange P 2021 {\em J. Phys. A: Math. Theor.\/} {\bf 54} 294001
	
	\bibitem{MajMouSabSch-2021}
	Majumdar S~N, Mounaix P, Sabhapandit S and Schehr G 2021 {\em
		arXiv:2110:01539\/}
	
	\bibitem{MerBoy-2021}
	Mercado-V\'{a}squez G and Boyer D 2021 {\em J. Phys. A: Math. Theor.\/} {\bf
		54} 444002
	
	\bibitem{SanDasNat-2021}
	Santra I, Das S and Nath S~K 2021 {\em J. Phys. A: Math. Theor.\/} {\bf 54}
	334001
	
	\bibitem{SchBre-2021}
	Schumm R~D and Bressloff P~C 2021 {\em J. Phys. A: Math. Theor.\/} {\bf 54}
	404004
	
	\bibitem{SinSanIom-2021}
	Singh R~K, Sandev T, Iomin A and Metzler R 2021 {\em J. Phys. A: Math.
		Theor.\/} {\bf 54} 404006
	
	\bibitem{Nav-2018}
	Navasqu\'es M 2018 {\em Phys. Rev. X\/} {\bf 8} 031008
	
	\bibitem{PerCarMag-2021}
	Perfetto G, Carollo F, Magoni M and Lesanovsky I 2021 {\em Phys. Rev. B\/} {\bf
		104} L180302
	
	\bibitem{WalBot-2021}
	Wald S and B{\"o}ttcher L 2021 {\em Phys. Rev. E\/} {\bf 103} 012122
	
	\bibitem{ReuUrbKla-2014}
	Reuveni S, Urbakh M and Klafter J 2014 {\em Proc. Natl. Acad. Sci. USA\/} {\bf
		111} 4391
	
	\bibitem{RotReuUrb-2015}
	Rotbart T, Reuveni S and Urbakh M 2015 {\em Phys. Rev. E\/} {\bf 92} 060101(R)
	
	\bibitem{RolLisSanGri-2016}
	Rold\'{a}n {\'E}, Lisica A, S\'{a}nchez-Taltavull D and Grill S~W 2016 {\em
		Phys. Rev. E\/} {\bf 93} 062411
	
	\bibitem{RobHadUrb-2019}
	Robin T, Hadany L and Urbakh M 2019 {\em Phys. Rev. E\/} {\bf 99} 052119
	
	\bibitem{LubSinZuc-1993}
	Luby M, Sinclair A and Zuckerman D 1993 {\em Inf. Proc. Lett.\/} {\bf 47} 173
	
	\bibitem{MonZec-2002}
	Montanari A and Zecchina R 2002 {\em Phys. Rev. Lett.\/} {\bf 88} 178701
	
	\bibitem{StoSanKoc-2021}
	Stojkoski V, Sandev T, Kocarev L and A P 2021 {\em Phys. Rev. E\/} {\bf 104}
	014121
	
	\bibitem{EvaMajSch-2020}
	Evans M~R, Majumdar S~N and Schehr G 2020 {\em J. Phys. A: Math. Theor.\/} {\bf
		53} 193001
	
	\bibitem{PalKosReu-2021}
	Pal A, Kostinski S and Reuveni S 2021 {\em J. Phys. A: Math. Theor.\/}  In
	press
	
	\bibitem{WanCheKan-2021}
	Wang W, Cherstvy A~G, Kantz H, Metzler R and Sokolov I~M 2021 {\em Phys. Rev.
		E\/} {\bf 104} 024105
	
	\bibitem{BodCheSok-2019}
	Bodrova A~S, Chechkin A~V and Sokolov I~M 2019 {\em Phys. Rev. E\/} {\bf 100}
	012119
	
	\bibitem{BodCheSok-2019-Renew}
	Bodrova A~S, Chechkin A~V and Sokolov I~M 2019 {\em Phys. Rev. E\/} {\bf 100}
	012120
	
	\bibitem{MonVil-2013}
	Montero M and Villarroel J 2013 {\em Phys. Rev. E\/} {\bf 87} 012116
	
	\bibitem{MenCam-2016}
	M\'{e}ndez V and Campos D 2016 {\em Phys. Rev. E\/} {\bf 93} 022106
	
	\bibitem{Shk-2017}
	Shkilev V~P 2017 {\em Phys. Rev. E\/} {\bf 96} 012126
	
	\bibitem{MonMasVil-2017}
	Montero M, Mas\'{o}-Puigdellosas A and Villarroel J 2017 {\em Eur. Phys. J.
		B\/} {\bf 90} 176
	
	\bibitem{BodSok-2020-CTRWRes}
	Bodrova A~S and Sokolov I~M 2020 {\em Phys. Rev. E\/} {\bf 101} 062117
	
	\bibitem{KusMajSabSch-2014}
	Ku\'{s}mierz {\L}, Majumdar S~N, Sabhapandit S and Schehr G 2014 {\em Phys.
		Rev. Lett.\/} {\bf 113} 220602
	
	\bibitem{KusGod-2015}
	Ku\'{s}mierz {\L} and Gudowska-Nowak E 2015 {\em Phys. Rev. E\/} {\bf 92}
	052127
	
	\bibitem{ZhoXuDen-2021}
	Zhou T, Xu P and Deng W 2021 {\em Phys. Rev. E\/} {\bf 104} 054124
	
	\bibitem{EvaMaj-2018}
	Evans M~R and Majumdar S~N 2018 {\em J. Phys. A: Math. Theor.\/} {\bf 51}
	475003
	
	\bibitem{Mas-2019}
	Masoliver J 2019 {\em Phys. Rev. E\/} {\bf 99} 012121
	
	\bibitem{RAD-2021}
	Radice M 2021 {\em Phys. Rev. E\/} {\bf 104} 044126
	
	\bibitem{PerCarLes-2021}
	Perfetto G, Carollo F and Lesanovsky I 2021 {\em arXiv:2112.05078\/}
	
	\bibitem{MagCarPer-2022}
	Magoni M, Carollo F, Perfetto G and Lesanovsky I 2015 {\em arXiv:2202.12655\/}
	
	\bibitem{MukSenMaj-2018}
	Mukherjee B, Sengupta K and Majumdar S~N 2018 {\em Phys. Rev. B\/} {\bf 98}
	104309
	
	\bibitem{TalPalSek-2020}
	Tal-Friedman O, Pal A, Sekhon A, Reuveni S and Roichman Y 2020 {\em J. Phys.
		Chem. Lett.\/} {\bf 11} 7350
	
	\bibitem{BesBovPet-2020}
	Besga B, Bovon A, Petrosyan A, Majumdar S~N and Ciliberto S 2020 {\em Phys.
		Rev. Res.\/} {\bf 2} 032029(R)
	
	\bibitem{GupPlaKun-2021}
	Gupta D, Plata C~A, Kundu A and Pal A 2021 {\em J. Phys. A: Math. Theor.\/}
	{\bf 54} 025003
	
	\bibitem{MasCamMen-2019}
	Mas\'o-Puigdellosas A, Campos D and M\'endez V 2019 {\em Front. Phys.\/} {\bf
		7} 112
	
	\bibitem{MasCamMen-2019JSTAT}
	Mas\'o-Puigdellosas A, Campos D and M\'endez V 2019 {\em J. Stat. Mech.\/} {\bf
		033201}
	
	\bibitem{PalKusReu-2019}
	Pal A, Ku\'{s}mierz {\L} and Reuveni S 2019 {\em New J. Phys.\/} {\bf 21}
	113024
	
	\bibitem{PalKusReu-2019PRE}
	Pal A, Ku\'{s}mierz {\L} and Reuveni S 2019 {\em Phys. Rev. E\/} {\bf 100}
	040101(R)
	
	\bibitem{MasCamMen-2019PRE}
	Mas\'o-Puigdellosas A, Campos D and M\'endez V 2019 {\em Phys. Rev. E\/} {\bf
		100} 042104
	
	\bibitem{PalKusReu-2020}
	Pal A, Ku\'{s}mierz {\L} and Reuveni S 2020 {\em Phys. Rev. Res.\/} {\bf 2}
	043174
	
	\bibitem{BodSok-2020-BrResI}
	Bodrova A~S and Sokolov I~M 2020 {\em Phys. Rev. E\/} {\bf 101} 052130
	
	\bibitem{EulMet-2016}
	Eule S and Metzger J~J 2016 {\em New. J. Phys.\/} {\bf 18} 033006
	
	\bibitem{PalKunEva-2016}
	Pal A, Kundu A and Evans M~R 2016 {\em J. Phys. A: Math. Theor.\/} {\bf 49}
	225001
	
	\bibitem{PalReu-2017}
	Pal A and Reuveni S 2017 {\em Phys. Rev. Lett.\/} {\bf 118} 030603
	
	\bibitem{CheSok-2018}
	Chechkin A and Sokolov I~M 2018 {\em Phys. Rev. Lett.\/} {\bf 121} 050601
	
	\bibitem{Abr-Steg}
	Abramowitz M and Stegun I~A 1974 {\em Handbook of mathematical functions\/}
	(New York: Dover)
	
	\bibitem{Ros}
	Ross S~M 2020 {\em Introduction to Probability and Statistics for Engineers and
		Scientists\/} (Amsterdam: Elsevier)
	
	\bibitem{Red}
	Redner S 2001 {\em A guide to first-passage processes\/} (Cambridge: Cambridge
	University Press)
	
	\bibitem{GraRyz}
	Gradshteyn I~S and Ryzhik I~M 2007 {\em Table of integrals, series, and
		products\/} (Amsterdam: Elsevier academic press)
	
	\bibitem{NIST}
	Olver F~W~J, Lozier D~W, Boisvert R~F and Clark C~W (eds) 2010 {\em NIST
		Handbook of Mathematical Functions\/} (Cambridge: Cambridge University Press)
	
\end{thebibliography}
\end{document}